\documentclass{article}

\usepackage{graphicx}  % standard LaTeX graphics tool;
\usepackage{amsmath}   % For getting proper math figns
\usepackage[compress]{cite}
\usepackage{amssymb}   % Used for getting various symbols eg. gtrsim, lesssim etc.
\usepackage{bm} % For bold math (esp of lower greek letters)
\usepackage{dcolumn}% Align table columns on decimal point
\usepackage{color}
\usepackage{mathrsfs}
\usepackage{amsfonts}
\usepackage{varioref}
\usepackage{csquotes}
\usepackage{hyphenat}
\usepackage{bigints}
\usepackage{float}
\usepackage[caption=false]{subfig}
\RequirePackage[colorlinks,citecolor=blue,urlcolor=magenta,linkcolor=blue]{hyperref}

\def \MP {Mathisson-Papapetrou}
\def \MPD {Mathisson-Papapetrou-Dixon}
\def \SSC {spin supplementary condition}
\def \TD {Tulczyjew-Dixon}

\def \S {Schwarzschild}
\def \SV {Sochenev}
\title{Resonant orbits for a spinning particle in Kerr spacetime}
\author{Sajal Mukherjee\footnote{Current affiliation: The Inter-University Centre for Astronomy and Astrophysics (IUCAA), Post Bag 4, Pune-411007, India; email: sajal@iucaa.in}$~^{1}$~~and~~Santanu Tripathy \footnote{st12rs048@iiserkol.ac.in}$~^{1,2}$\\
{$~^{1}$ \small{Department of Physical Sciences, IISER-Kolkata, Mohanpur-741246, India.}}\\
{$~^{2}$ \small{Center of Excellence in Space Sciences India, IISER-Kolkata, Mohanpur-741246, India.}}}

\begin{document}
\maketitle
\begin{abstract}
\noindent
In the present article, we study the orbital resonance corresponds to an extended object approximated up to the dipole order term in Kerr spacetime. We start with the \MP~equations under the linear spin approximation and primarily concentrate on two particular events. First, when the orbits are nearly circular and executing a small oscillation about the equatorial plane and second, a generic trajectory confined on the equatorial plane. While in the first case, all the three fundamental frequencies, namely, radial $\Omega_r$, angular $\Omega_{\theta}$, azimuthal $\Omega_{\phi}$ can be commensurate with each others and give rise to the resonance phenomenon, the later is only accompanied with the resonance between $\Omega_r$ and $\Omega_{\phi}$ as we set $\theta=\pi/2$. We provide a detail derivation in locating the prograde resonant orbits in either of these cases and also study the role played by the spin of the black hole. The implications related to spin-spin interactions between the object and black hole are also demonstrated. 
\end{abstract}
%%%%%%%%%%%%%%%%%%%%%%%%%%%%%%%%%%%%%%%%%%%%%%%%%%%%%%%%%%%%%%%%%%%%%%%%%%%%%%%%%%%%%%%%%%%%%%
\section{Introduction}
In the strong field regime where the effects of gravity are inescapable, the motion of test particles can attribute to many nontrivial consequences such as precession, spiraling of orbits, resonance, chaos and so on \cite{einstein1915erklarung,meaning,will1993theory,Brink:2013nna,Brink:2015roa,Suzuki:1996gm}. Even if these investigations related to the trajectories of test particles are not only present day concerns rather contain with extensive literature spanned over last several years \cite{Bardeen:1972fi,o1995geometry}, there are relevant studies with possible observational aspects yet to be explored \cite{Levin:2008mq,Grossman:2011ps}. Regarding the observational implications, there are proposed mission such as The Laser Interferometer Space Antenna ( LISA ) which aims to probe extreme mass ratio binaries in near future as well \cite{danzmann2011lisa,sopuerta2012probing,Armano:2016bkm}. However, in any attempt to explain the phenomenon involves incomparable masses would require to grasp the theoretical structure of particle trajectories as well. While in many cases, it is appropriate to consider that the lighter companion as a point test particle that moves along a geodesic trajectory, the addition of higher order moments may introduce larger accuracy in estimating any measurable quantities. These higher order moments can originate from the non-trivial internal structure of the body represented by its energy momentum tensor $T^{\mu \nu}$ while their dynamics can be derived from the conservation equation $\nabla_{\mu}T^{\mu \nu}=0$. By expanding $T^{\mu \nu}$ about a reference point located inside the object, say $z_{\mu}(t)$, the zeroth order term of the conservation equation would produce the geodesic equation while the first order introduces the \MP~equations accordingly \cite{Mathisson:1937zz,Papapetrou:1951pa}. The next order contribution arise from the quadrupole moment and would be given by the \MPD~equations \cite{dixon1970dynamics,dixon1970dynamics2,dixon1974dynamics}. However, in the present article, we only consider the first order correction to the geodesic equations as introduced by the nonzero spin or dipole moment of the particle $S^{\mu \nu}$ given as \cite{Papapetrou:1951pa}
%%%%%%%%%%%%%%%%%%%%%%%%%%%%%%%
\begin{equation}
S^{\mu \nu}=\int{r^{\mu}\bold{T}^{0 \nu}d^3x}-\int{r^{\nu}\bold{T}^{0 \mu}d^3x},
\label{eq:spin_tensor_definition}
\end{equation}
%%%%%%%%%%%%%%%%%%%%%%%%%%%%%%%
with both $\mu$ and $\nu$ running from 0 to 3. The scaled energy-momentum tensor $\bold{T}^{0 \mu}=\sqrt{-g}~T^{0 \mu}$ with $g$ being the determinant of the metric, is integrated over the spacelike hypersurface represented by $t=\text{constant}$ slices. Dipole moment is computed about a reference point $z_{\mu}(t)$ while the distance between $z_{\mu}(t)$ and any given mass point is given by $r^{\mu}$ which contains no time-component according to the definition.

From a theoretical standpoint, the astrophysical objects are likely to have a finite size and therefore it is appropriate to consider additional moments such as dipole, quadrupole and so on. While particle with only mass monopole are named as monopole particle, the addition of dipole and quadrupole moment would introduce pole-dipole, i.e., a spinning particle, and pole-dipole-quadrupole particle respectively. However, in the present context we shall only discuss some aspects of pole-dipole or spinning particle and not consider the quadrupole moment. The motion of a spinning particle constitutes an interesting problem and studied extensively in literature \cite{papapetrou1955pol,Hojman:1976kn,Jefremov:2015gza,Plyatsko:2013xza}. With the pioneering work by Papapetrou in 1951 \cite{Papapetrou:1951pa}, the early contributions from Mathisson \cite{Mathisson:1937zz}, Ehlers and Rudolph \cite{ehlers1977dynamics} and Dixon \cite{dixon1964covariant} are worth to mention. In recent times, there are also significant studies regarding several aspects of spinning particle and their nontrivial features \cite{PhysRevD.92.064032,Kyrian:2007zz,Costa:2011zn,Costa:2017kdr,Lukes-Gerakopoulos:2014dma,Tanaka:1996ht,corinaldesi2003spinning,Saijo:1998mn,Mukherjee:2018zug,Mukherjee:2018bsn,Mukherjee:2018kju,Deriglazov:2017jub,Deriglazov:2018vwa}. For excellent overview on the subject, we refer our readers Refs.\cite{Semerak:1999qc,Steinhoff:2015ksa,Semerak:2015lnv}. In one way, it emerges as an engaging theoretical implication with possible usefulness in explaining motion of objects with non-trivial internal structure and also pave the way for extreme mass ratio inspiral in general \cite{Babak:2014kqa}.

In the present article, we aim to study the resonant orbits for a spinning particle in the Kerr background and for computational convenience, we restrict our discussions on two major aspects: first, the resonance between different small oscillation frequencies correspond to nearly circular orbits close to the equatorial plane and second, the resonance between radial and azimuthal frequencies considering the trajectories are completely confined on the equatorial plane of the black hole. In the former, it is possible to encounter the resonance in-between radial $\Omega_{r}$, angular $\Omega_{\theta}$ as well as azimuthal $\Omega_{\phi}$ frequencies while in the later, it is only between $\Omega_{r}$ and $\Omega_{\phi}$. Out of these three frequencies, the radial and angular are known as libration frequencies, while the later $\Omega_{\phi}$ can be thought of as a rotational frequency. The correspondence between these frequencies would lead to the occurrence of resonance phenomenon in astrophysical scenarios. In particular, whenever the ratio of any of these frequencies become rational fraction, the orbits become resonant. For example, with the $r\phi$ resonance, the principle equation is given as $\Omega_r/\Omega_{\phi}=n/m$, where both $n$ and $m$ are rational number with no common divisor.

With the above motivation, let us now introduce the physical notion of resonance in astrophysical scenarios and also highlight its importance in the present context as well. Apart from astrophysical settings, the events like resonance are usually encountered in vibrations of strings and coupled oscillations \cite{bajaj1988physics}. For example, in case of a forced vibrating oscillator, the resonance becomes dominant whenever the natural frequency of the system becomes nearly equal to the frequency of the forced oscillation. Similar to this, the weakly coupled oscillators can also undergo resonance phases whenever their individual frequencies are comparable to each other. On the other hand, any external perturbation may introduce non-integrability in the Hamiltonian and the system may encounter chaos. However, for a sufficiently small perturbation, the system may remain nearly integrable and the notion of separability constant (also known as Carter constant in the black hole spacetime) may still approximately exist \cite{Carter:1969zz,Rosquist:2007uw,Mukherjee:2015oaa}. In the case of a spinning particle with \TD~supplementary condition, the Carter-like constant can be defined as far as the $\mathcal{O}(S^2)$ terms are neglected \cite{Witzany:2019dii,Witzany:2019nml,Gibbons:1993ap,Tanaka:1996ht}. However, for other supplementary conditions, the spin of the particle may introduce chaos in the system \cite{Kunst:2015tla,Lukes-Gerakopoulos:2016bup}. Nonetheless, as the present analysis only consider the \TD~condition, the system can be considered integrable within the linear spin approximation. This would allow us to obtain the resonant orbits similar to the case of geodesic trajectories as given in Ref. \cite{Brink:2015roa}. It is of significant interest to locate the resonant orbits as it is likely that these orbits will witness the breakdown of phase space tori for the first time if the system undergoes any chaos \cite{Zelenka:2019nyp,arnold1963proof,arnold1963small,moser1973stable}.

Earlier in literature, several implications related to resonance activities are addressed in various aspects \cite{Berry:2016bit,vandeMeent:2014raa,Hirata:2010xn}. As already mentioned, the orbital dynamics in Kerr depends on three fundamental frequencies relating three spatial components and therefore, there could exists different types of resonant orbits. The $r\theta$ resonant orbits are studied in Ref. \cite{Brink:2015roa} whereas the $\theta\phi$ and $r\phi$ resonances are addressed in Ref. \cite{Hirata:2010xn} and Ref. \cite{vandeMeent:2014raa} 
respectively. In Refs. \cite{Hirata:2010xn,vandeMeent:2014raa}, the black hole kicks are also studied for an extreme mass ratio inspiral considering the lighter companion follows a geodesic orbit. For an 
investigation related to the resonance in between the spin precession 
frequencies and orbital frequencies of a spinning particle, we refer our readers Ref. \cite{Ruangsri:2015cvg} for a better understanding. However, as far as the analysis in the present article is concerned, we would only consider the resonance between orbital fundamental frequencies of a spinning particle and the effect 
of the spin precession frequency is excluded.

The rest of the paper is organized as follows. In section-(\ref{sec:motion_spinning}), the basic governing equations of a spinning particle are introduced along with various implications such as \SSC~and conserved quantities. These equations are exactly solved on the equatorial plane of the Kerr black hole and the effective radial potential is also obtained for further calculations. Following this in section-(\ref{sec:resonance}), we introduce the machinery to obtain the fundamental frequencies and orbital resonance for any trajectory in black hole's spacetime. In particular, we studied two different events corresponding to a spinning object: first, the small oscillation frequencies for a particle orbiting in nearly circular orbits close to the equatorial orbits and second, a general motion confined on $\theta=\pi/2=\text{constant}$ plane. Section-(\ref{sec:nearly}) is devoted to explore the resonance phenomenon related to the first case while in section-(\ref{sec:rphi}), the $r\phi$ resonance are elaborately studied. In the later, the approximation technique namely the Sochnev method is employed to evaluate the integrals on the equatorial plane of the black hole. We conclude the article with a brief remark in section-(\ref{sec:remarks}).
%%%%%%%%%%%%%%%%%%%%%%%%%%%%%%%%%%%%%%%%%%%%%%%%%%%%%%%%%%%%%%%%%%%%%%%%%%%%%%%%%%%%%%%%%%%%%%%%e
\section{Motion of a spinning particle in gravitational field} \label{sec:motion_spinning}
The motion of a spinning particle is described by the \MP~equations
%%%%%%%%%%%%%%%%%%%%%%%%%%%%
\begin{equation}
\dfrac{DP^a}{d\tau} = -\dfrac{1}{2}R^a_{~b c d}\mathcal{U}^b S^{c d}, \qquad \dfrac{DS^{a b}}{d\tau} = P^a \mathcal{U}^b-P^b \mathcal{U}^a,
\label{eq:MP_eq_spin}
\end{equation}
%%%%%%%%%%%%%%%%%%%%%%%%%%%%
with $P^a$ and $\mathcal{U}^a$ correspond to the four momentum and four velocity of the particle respectively, $S^{ab}$ is the spin tensor of the extended object relating the dipole moment and $R^a_{~bcd}$ has the usual meaning of Riemann curvature tensors correspond to the background geometry. Furthermore, it should be emphasized that the above set contains total 10 equations while we have $14$ (6 from the spin, 4 from the  four velocity and 4 from the four momentum) unknowns to solve and therefore, additional conditions are required. These are called \SSC~and in the present context, we shall employ the \TD~condition $S^{ij}P_{j}=0$ to solve \MP~equations around the Kerr black hole \cite{tulczyjew1959motion}. While a supplementary condition only gives three independent constraint, the fourth comes from the choice of the time parametrization \cite{Lukes-Gerakopoulos:2017cru}. In the present article, we have employed the parametrization introduced in Ref. \cite{ehlers1977dynamics} and later employed in Refs. \cite{Saijo:1998mn,Tanaka:1996ht}. This is given as $P^{i}\mathcal{U}_{i}=-\mu$, where $\mu$ is the conserved mass as given by $P^{i}P_{i}=-\mu^2$. Let us now introduce the notion of spin vector $S^{i}$ for simplified computations and it is related with the spin tensor as follows \cite{Tanaka:1996ht,Saijo:1998mn}
%%%%%%%%%%%%%%%%%%%%%%%%%%%
\begin{equation}
S^{i}=\dfrac{1}{\mu \sqrt{-g}}\epsilon^{ijkl}P_{j}S_{kl},
\end{equation}
%%%%%%%%%%%%%%%%%%%%%%%%%%%
where $g$ is the determinant of the background metric. For the particle confined on the equatorial plane, it is convenient to assume that the spin vector follows $S^{i}=(0,0,S^{\theta},0)$ which indicates that the spin is either parallel or anti-parallel to the black hole's rotational axis.

In addition, the symmetries associated with the background geometry will give rise to the conserved quantities such as energy and momentum. The compact expression for any conserved quantity $\mathcal{C}$ related to a killing vector field $\mathcal{K}^{i}$ can be written as
%%%%%%%%%%%%%%%%%%%%%%%%%%%%
\begin{equation}
\mathcal{C}=\mathcal{K}^{i}P_{i}-\dfrac{1}{2}S^{ij}\mathcal{K}_{i;j},
\end{equation}
%%%%%%%%%%%%%%%%%%%%%%%%%%%%  
where the contribution from the spin can be easily spotted \cite{rudiger1981conserved,rudiger1983conserved}. It should also be noted that the total spin per mass of the particle, i.e., $S^2= (S^{i}S_{i})/\mu^2$ is conserved along the trajectory of the particle as it can be established from the \TD~supplementary condition. With all these machinery, one should be able to solve the \MP~equations and also establish a relation between four velocity and four momentum of the particle. In the present context, we only state the final results corresponding to the components of four velocity of the particle while the full derivations can be found in Ref. \cite{Saijo:1998mn}. On the equatorial plane, these equations are given as
%%%%%%%%%%%%%%%%%%%%%%%%%%%%%%%%
\begin{eqnarray}
(\Sigma_s\Lambda_s\mathcal{U}^1)^2 &=& R_{\rm s}= P_s^2-\Delta \left( \dfrac{ \Sigma_s^2}{r^2} +\left\{J_z-(a+S)E\right\}^2\right) \,, \nonumber \\
(\Sigma_s\Lambda_s\mathcal{U}^0) & = & a\left(1+\dfrac{3 S^2}{r \Sigma_s }\right)\left \{J_z-(a+S)E \right\}+\dfrac{r^2+a^2}{\Delta}P_s\,, \nonumber \\
(\Sigma_s\Lambda_s\mathcal{U}^{3}) & = & \left(1+\dfrac{3 S^2}{r \Sigma_s }\right)\left \{J_z-(a+S)E \right \}+\dfrac{a}{\Delta} P_s \,,
\label{eq:eom}
\end{eqnarray}
%%%%%%%%%%%%%%%%%%%%%%%%%%%%%%%%
where $a$ and $M$ are angular momentum and mass parameter of the black hole respectively; and $E$ and $J_{z}$ are given as energy and momentum per mass of the particle. In the above, we assume $(0,1,2,3)$ are $(t,r,\theta,\phi)$ respectively. The expressions for $P_s$ and $\Sigma_s$ are given as
%%%%%%%%%%%%%%%%%%%%%%%%%%%%%%
\begin{eqnarray}
P_s &= &   E \left(r^2+a^2 +a S+\frac{a S M}{r}\right)-\left(a+ \frac{M S}{r}\right)J_z ,\, \nonumber \\
\Sigma_s &=& r^2  \left(1-M S^2/r^3 \right) \quad ; \quad \Delta=r^2+a^2-2 M r ,\, \nonumber \\
\Lambda_s &=& 1- \frac{3 M S^2 r}{\Sigma_s^3}\left\{J_z-(a+S)E\right\}^2.
\label{Eq:Ps}
\end{eqnarray}
%%%%%%%%%%%%%%%%%%%%%%%%%%%%%%%%
For $S>0$, the spin is parallel to z-axis, and anti-parallel for $S<0$. In addition, to guarantee that the timelike constraint $\mathcal{U}^i\mathcal{U}_i<0$ is always valid, we have the following condition to hold
%%%%%%%%%%%%%%%%%%%%%%%%%%%
\begin{equation}
r^5(1-MS^2/r^3)^4-3 M S^2(2+MS^2/r^3)\left\{J_z-E(a+S)\right\}^2>0.
\end{equation}
%%%%%%%%%%%%%%%%%%%%%%%%%%% 
Needless to say that the above condition further constraint the motion of the particle and especially close to the horizon, it becomes more dominant \cite{Mukherjee:2018kju}. 
%%%%%%%%%%%%%%%%%%%%%%%%%%%%%%%%%%%%%%%%%%%%%%%%%%%%%%%%%%%%%%%%%%%%%%%%%%%%
\section{Resonance phenomenon and fundamental frequencies} \label{sec:resonance}
The general condition for resonance can be written in terms of the fundamental frequencies, such that
%%%%%%%%%%%%%%%%%%%%%%%%%%%%%%%%%%%%
\begin{equation}
\alpha \Omega_{r}+\beta \Omega_{\theta}+\gamma \Omega_{\phi}=0 ,
\label{eq:resonance_condition_first}
\end{equation}
%%%%%%%%%%%%%%%%%%%%%%%%%%%%%%%%%%%%
where $\alpha,\beta$, and $\gamma$ are rational numbers without any common divisors. For a limiting case on the equatorial plane, $\Omega_{r}$ and $\Omega_{\phi}$ would be of the particular interest, while $\Omega_{\theta}$ has no meaning. On the other hand, for circular orbits, $\Omega_{r}$ is not particularly relevant and $\Omega_{\theta}$ and $\Omega_{\phi}$ play the key role in the orbital dynamics. In the present context, we shall be interested in following investigations
%%%%%%%%%%%%%%%%%%%%%%%%%%%%%%%%%%%
\begin{itemize}
\item The quasi periodic oscillations for spinning particles and the locations of resonant orbits from the condition given by Eq. (\ref{eq:resonance_condition_first}).
\item Location of resonant orbits on the equatorial plane following the condition, $\alpha \Omega_r+\gamma \Omega_{\phi}=0$ which is the $r\phi$ resonance. 
\end{itemize}
%%%%%%%%%%%%%%%%%%%%%%%%%%%%%%%%%%%
In the first case, we consider that the particle is following nearly circular orbit, and also executing small oscillation about the equatorial plane. The detail calculations to obtain the fundamental frequencies are given  by Hinderer et. al. in Ref. \cite{Hinderer:2013uwa}, and we explicitly employ them in our purpose. We study the resonance phenomenon related to these frequencies in section-(\ref{sec:nearly}). However, as already mentioned, the present analysis is valid only at $\mathcal{O}(S)$, i.e., $|S| \ll 1$, which is 
exactly the appropriate regime to describe an extreme mass ratio system through the \MP~equaitons \cite{dixon1964covariant}.

%%%%%%%%%%%%%%%%%%%%%%%%%%%%%%%%%%%%%%%%%%%%%%%%%%%%
In the case of a generic elliptical orbit, we may introduce the following expressions for radial and azimuthal frequencies 
%%%%%%%%%%%%%%%%%%%%%%%%%%%%%%%%
\begin{equation}
\Omega_{r}=\dfrac{2\pi}{T_r}, \quad \text{and}\quad \Omega_{\phi}=\dfrac{\Delta \phi}{T_r},
\end{equation}
%%%%%%%%%%%%%%%%%%%%%%%%%%%%%%%% 
with, $T_r=2 \bigintsss_{r_a}^{r_p}{\biggl(\dfrac{dt}{dr}\biggr)dr}$ and $\Delta {\phi}=2 \bigintsss_{r_a}^{r_p}{\biggl(\dfrac{d\phi}{dr} \biggr)dr}$, $r_a$ and $r_p$ are the apastron and periastron radii respectively \cite{Cutler:1994pb,Kunst:2015hsa}. In this case, the $r \phi$ resonance would be governed by the equation
%%%%%%%%%%%%%%%%%%%%%%%%%%%%%%%%%
\begin{equation}
m \Omega_{r}-n \Omega_{\phi}=0, 
\end{equation}
%%%%%%%%%%%%%%%%%%%%%%%%%%%%%%%%%
with $m$ and $n$ are two integers constrained as $n \leq m$. More conveniently, the above equation can be written as
%%%%%%%%%%%%%%%%%%%%%%%%%%%%%%%%%
\begin{equation}
m \pi-n \int^{r_a}_{r_p}{\dfrac{d\phi}{dr}dr}=0.
\label{eq:primary}
\end{equation}
%%%%%%%%%%%%%%%%%%%%%%%%%%%%%%%%%
The second term in the above expression can be written as
%%%%%%%%%%%%%%%%%%%%%%%%%%%%%%%%%%%%%%
\begin{equation}
\dfrac{d\phi}{dr}=\Big(\dfrac{d\phi}{d\tau}\Big)\Big(\dfrac{dr}{d\tau}\Big)^{-1},
\label{eq:dphidrfst}
\end{equation}
%%%%%%%%%%%%%%%%%%%%%%%%%%%%%%%%%%%%%%
\noindent
with $\tau$ being the proper time. By comparing with the parametrization given in section-(\ref{sec:motion_spinning}), there seems to exist an apparent contradiction! However, it can be further explained as follows: given that the parametrization, $P^{i}\mathcal{U}_{i}=-\mu$, is valid, the affine parameter is given by $\sigma$ --- which may not be the proper time. Therefore, $\mathcal{U}^{i}$ is a tangent to the curve, but not the four velocity. In particular, the \MP~equations can now be written as
%%%%%%%%%%%%%%%%%
\begin{equation}
\dfrac{DP^a}{d\sigma} = -\dfrac{1}{2}R^a_{~b c d}\left(\dfrac{dx^b}{d\sigma}\right) S^{c d}, \qquad \dfrac{DS^{a b}}{d\sigma} = P^a \left(\dfrac{dx^b}{d\sigma}\right)-P^b \left(\dfrac{dx^a}{d\sigma}\right).
\end{equation}
%%%%%%%%%%%%%%%%%
However, with the relation,
%%%%%%%%%%%%%%%%
\begin{equation}
\dfrac{d}{d\sigma}=\dfrac{d\tau}{d\sigma}\dfrac{d}{d\tau},
\end{equation}
%%%%%%%%%%%%%%%%
being introduced, the above equation can be casted again as Eq. (\ref{eq:MP_eq_spin}), provided that $({d\tau}/{d\sigma})$ is nonzero. Therefore, the time parametrization given in section-(\ref{sec:motion_spinning}) can be employed to describe usual dynamics of a spinning particle. Moreover, in the present context, we have only considered terms up to linear order in spin, and therefore, the parametrization with respect to both $\tau$ and $\sigma$ would result the same. Coming back to Eq. (\ref{eq:dphidrfst}), it can be further simplified by introducing the orbit equations in terms of energy, angular momentum, spin and the radial coordinate $r$. However, as it is evident that the quantity $dr/d\tau$ is proportional to the radial potential which in fact, vanishes at the turning points $r_a$ and $r_p$. This constitutes a serious problem while integrating the function that blows up in the upper and lower limits. We shall use approximate technique, namely the Sochnov method, to deal with such scenarios.  
%%%%%%%%%%%%%%%%%%%%%%%%%%%%%%%%%%%%%%%%%%%%%%%%%%%%
%%%%%%%%%%%%%%%%%%%%%%%%%%%%%%%%%%%%%%%%%%%%%%%%%%%%%%%%%%%%%%%%%%%%%%%%%%%%%%%%%%%%%%%%%%%%%%%%%%%%%%
\section{Nearly circular and equatorial orbits and resonance conditions}\label{sec:nearly}
Let us now consider the nearly circular and equatorial orbits for spinning particles and determine the small oscillation frequencies accordingly. For this purpose we shall assume the spin vector has the form $S=(0,0,S^{\theta},0)$ and the four velocity can be written as $\mathcal{U}^{i}=(\mathcal{U}^0,0,0,\mathcal{U}^{3})$. In addition, we shall approximately write the frequencies up to the linear order in spin while neglecting all the higher order terms.
%%%%%%%%%%%%%%%%%%%%%%%%%%%%%%%%%%%%%%%%%%%%%%%%%%%%%%%%%%%%%%%%%%%%%%%%%%%%%%%%%%%%%%%%%%%%%%%%%%%%%%
\subsection{Fundamental frequencies $\Omega_{r}, \Omega_{\theta}$ and $\Omega_{\phi}$ }
%%%%%%%%%%%%%%%%%%%%%%%%%%%%%%%%%%%%%%%%%%%%%%%%%%%%%%%%%%%%%%%%%%%%%%%%%%%%%%%%%%%%%%%%%%
We now introduce the fundamental frequencies as given in Ref. \cite{Hinderer:2013uwa}. Given that we are only considering the terms linear in spin, the frequencies for both co-rotating and counter-rotating orbits take the following form:
%%%%%%%%%%%%%%%%%%%%%%%%%%
\begin{eqnarray}
\dfrac{\Omega^2_r}{\Omega^2_{\phi}} &= &\dfrac{(r_{\rm c}-6M)r_{\rm c}-3 a^2 \pm 8 a \sqrt{M r_{\rm c}}}{r_{\rm c}^2}+ \dfrac{6S (\pm  \sqrt{M r_{\rm c}}-a)\left[(r_{\rm c}-3M)\sqrt{r_{\rm c}} \pm 2 a \sqrt{M}\right]}{r_{\rm c}^{7/2}}+\mathcal{O}(S^2), \nonumber \label{eq:fun_r} \\
\\
\dfrac{\Omega_{\phi}}{\Omega_{\theta}} & = & \dfrac{r_{\rm c}}{\sqrt{3 a^2 \mp 4 a  \sqrt{M r_{\rm c}}+r^2_{\rm c}}}-\dfrac{3aS \left\{\pm 2 a \sqrt{M}+(r_{\rm c}-3M)\sqrt{r_{\rm c}}\right\}}{\sqrt{r_{\rm c}}\left\{3 a^2 \mp 4 a \sqrt{M r_{\rm c}}+r^2_{\rm c}\right\}^{3/2}}+\mathcal{O}(S^2),  \label{eq:fun_theta} \\
\Omega_{\phi} &=& \dfrac{\sqrt{M}}{r^{3/2}_{\rm c}\pm a \sqrt{M}},  \label{eq:fun_phi}
\end{eqnarray}
%%%%%%%%%%%%%%%%%%%%%%%%%%
where, $r_{\rm c}$ is the radius of the circular orbit about which the spinning particle is executing small oscillation; and upper and lower sign corresponds to co-rotating and counter rotating orbits respectively. In the next section, we employ the above frequencies to obtain the resonant orbits for a spinning particle. 
%%%%%%%%%%%%%%%%%%%%%%%%%%%%%%%%%%%%%%%%%%%%%%%%%%%
\subsection{Locations of the resonant orbits}
\begin{figure}[htp]
\subfloat[The $r\theta$ resonant orbits are shown as a function of the black hole's momentum while the spin of the particle takes different values.\label{fig:Resonance_01}]{%
  \includegraphics[height=6.8cm,width=.46\linewidth]{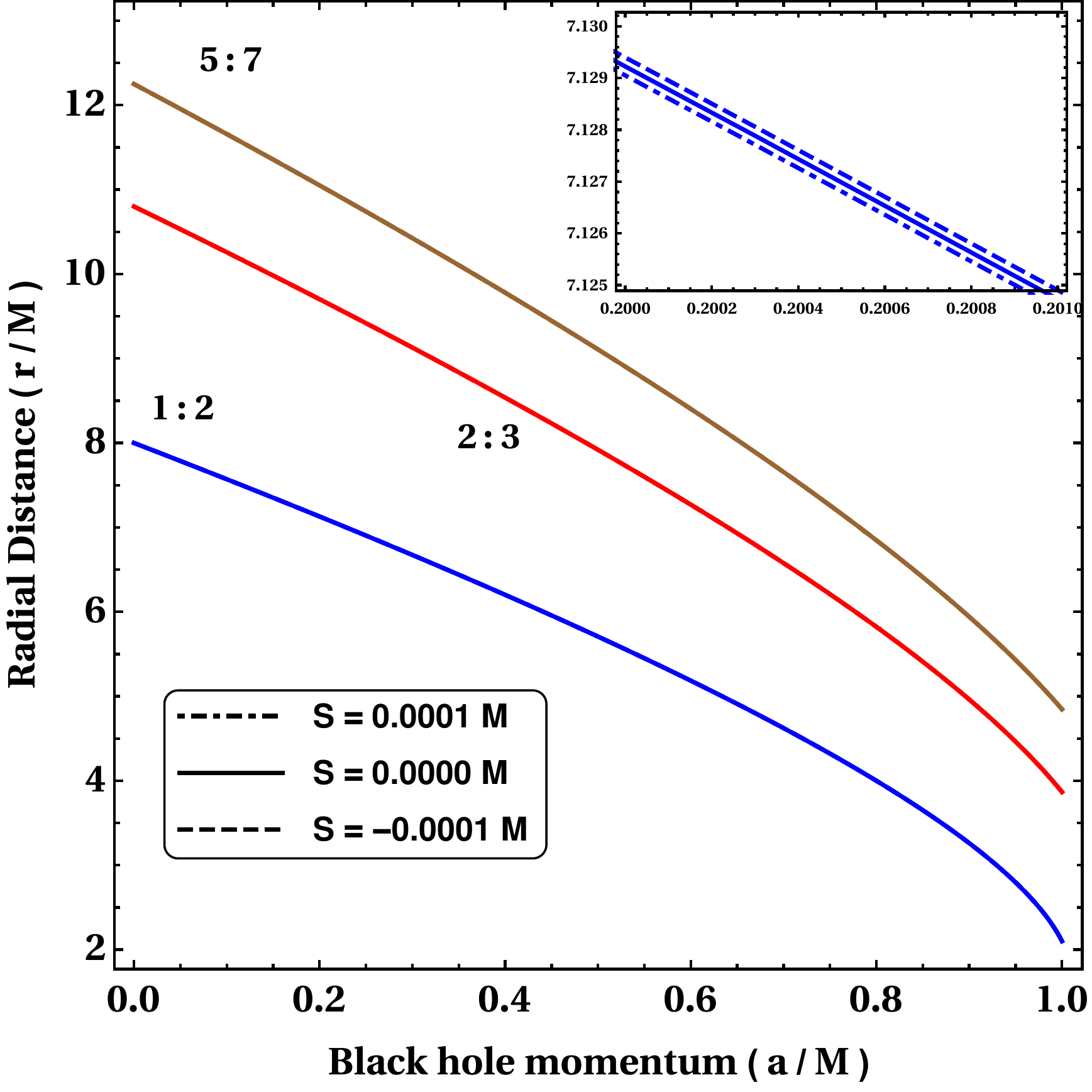}%
}\hfill
\subfloat[The above figure demonstrates the $r\phi$ resonance for different spin of the extended object. \label{fig:Resonance_02}]{%
  \includegraphics[height=6.8cm,width=.46\linewidth]{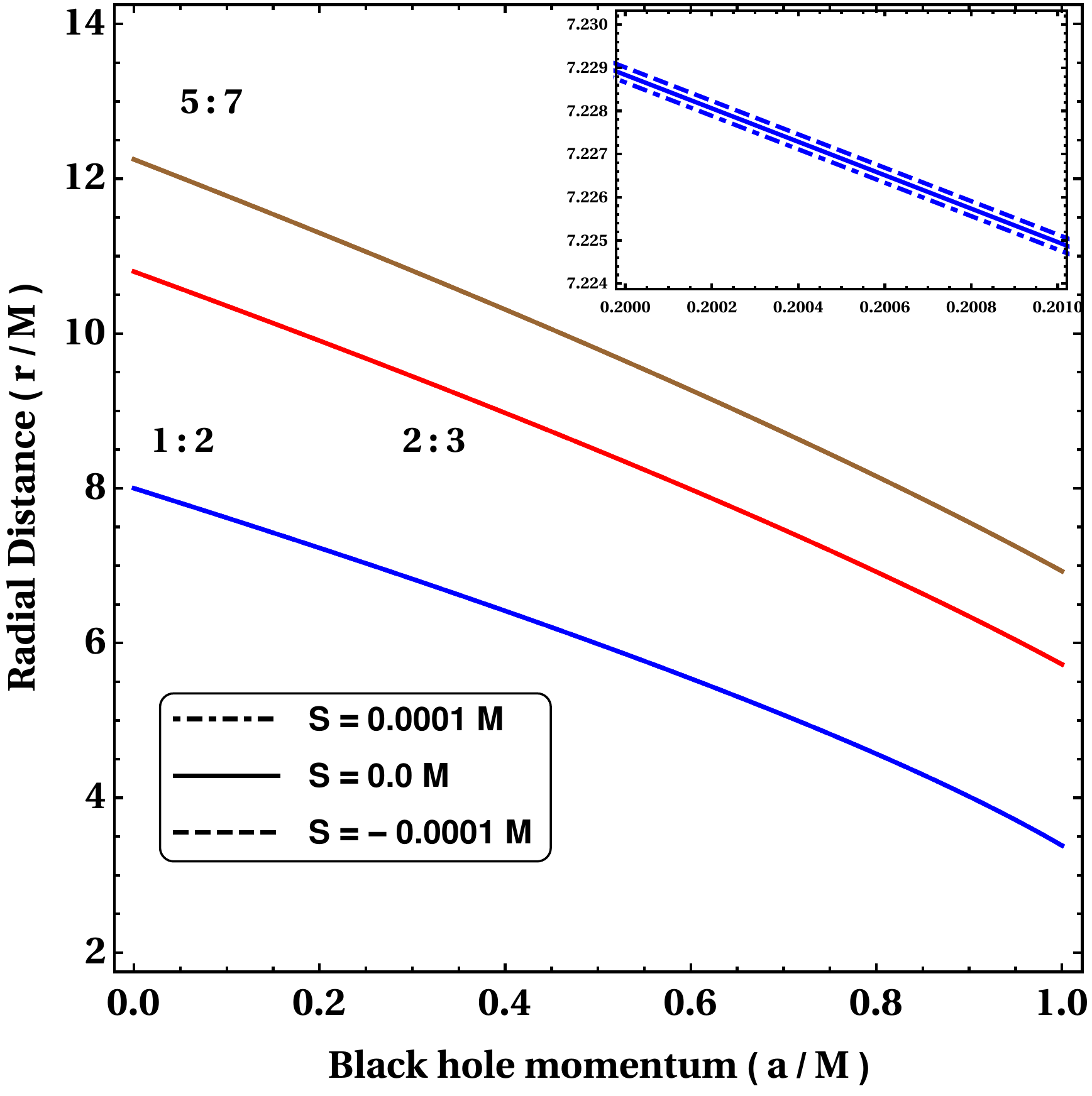}%
}\\
\centering
\subfloat[In the above figure, $\theta \phi$ resonance is depicted as a function of black hole's momentum.\label{fig:Resonance_02a}]{%
  \includegraphics[height=6.8cm,width=.46\linewidth]{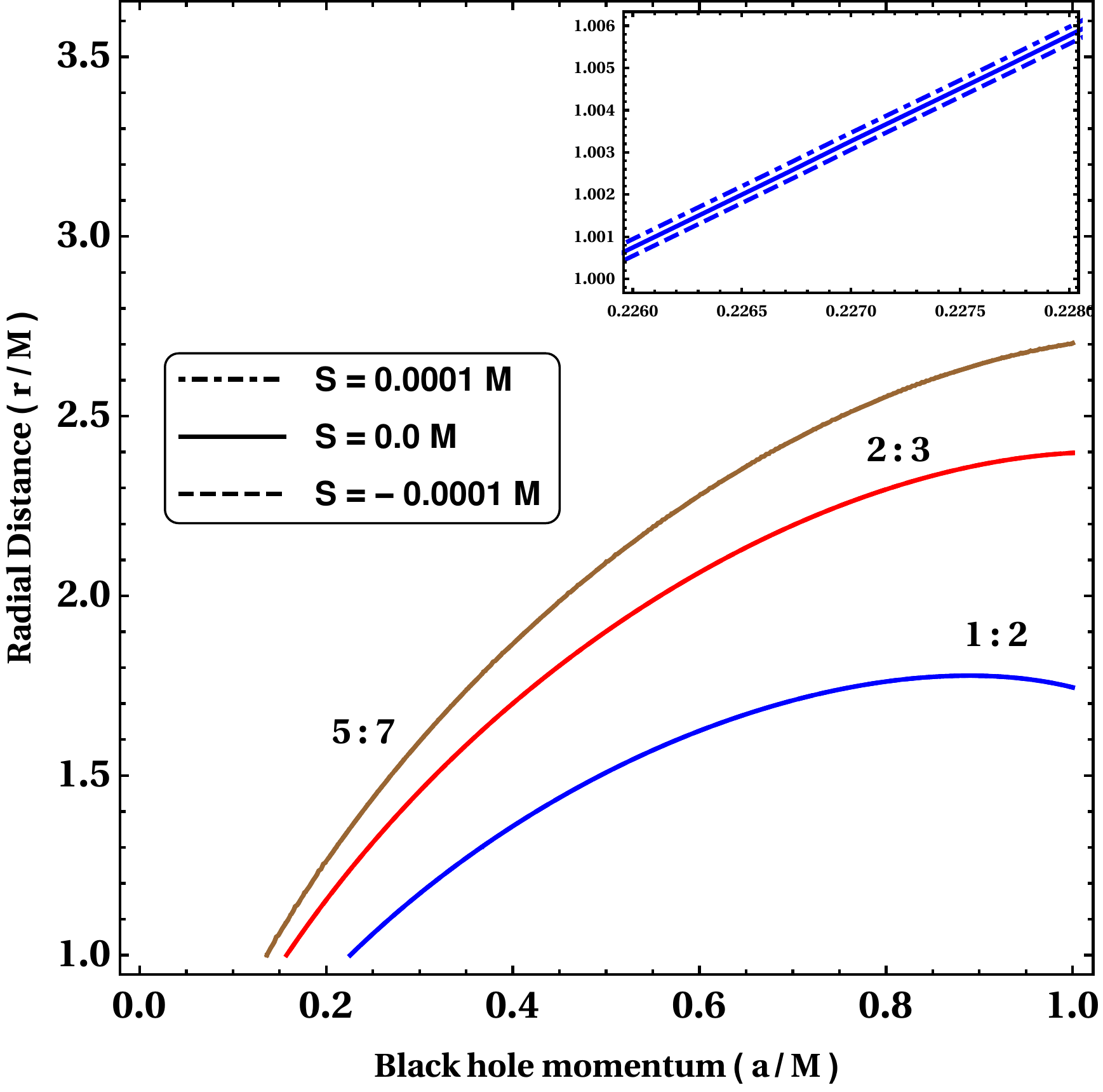}%
}\hfill
\caption{The above figures capture the location of co-rotating resonant orbits for different spin parameters and black hole's momentum. Here, we have used Eq. (\ref{eq:fun_r}), Eq. (\ref{eq:fun_theta}) and Eq. (\ref{eq:fun_phi}) to obtain the above figures. Distinguishing features related to various resonances are also depicted for different orders. The blue, red and brown colours are related to $1:2$, $2:3$ and $5:7$ resonance order respectively. In the inset of each plot, we have plotted the zoomed curve for $1:2$ resonance.}
\label{fig:Resonance_A}
\end{figure}

\begin{figure}[htp]
\subfloat[The above figure demonstrates the $r\theta$ resonance in counter-rotating orbits for different spin parameters.\label{fig:Resonance_01cnt}]{%
  \includegraphics[height=6.8cm,width=.46\linewidth]{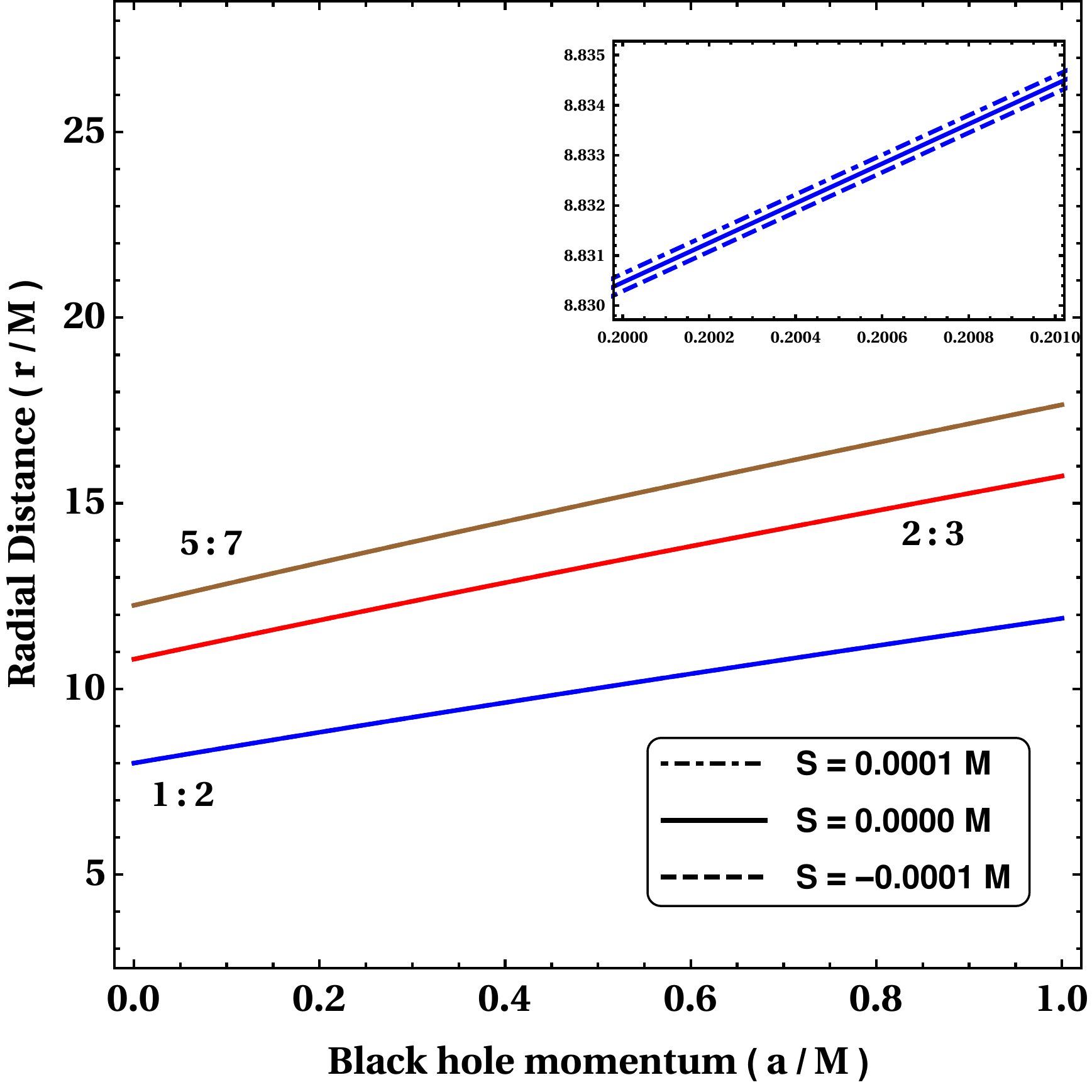}%
}\hfill
\subfloat[In the above, $r\phi$ resonance is shown as a function of the black hole's angular momentum. \label{fig:Resonance_02cnt}]{%
  \includegraphics[height=6.8cm,width=.46\linewidth]{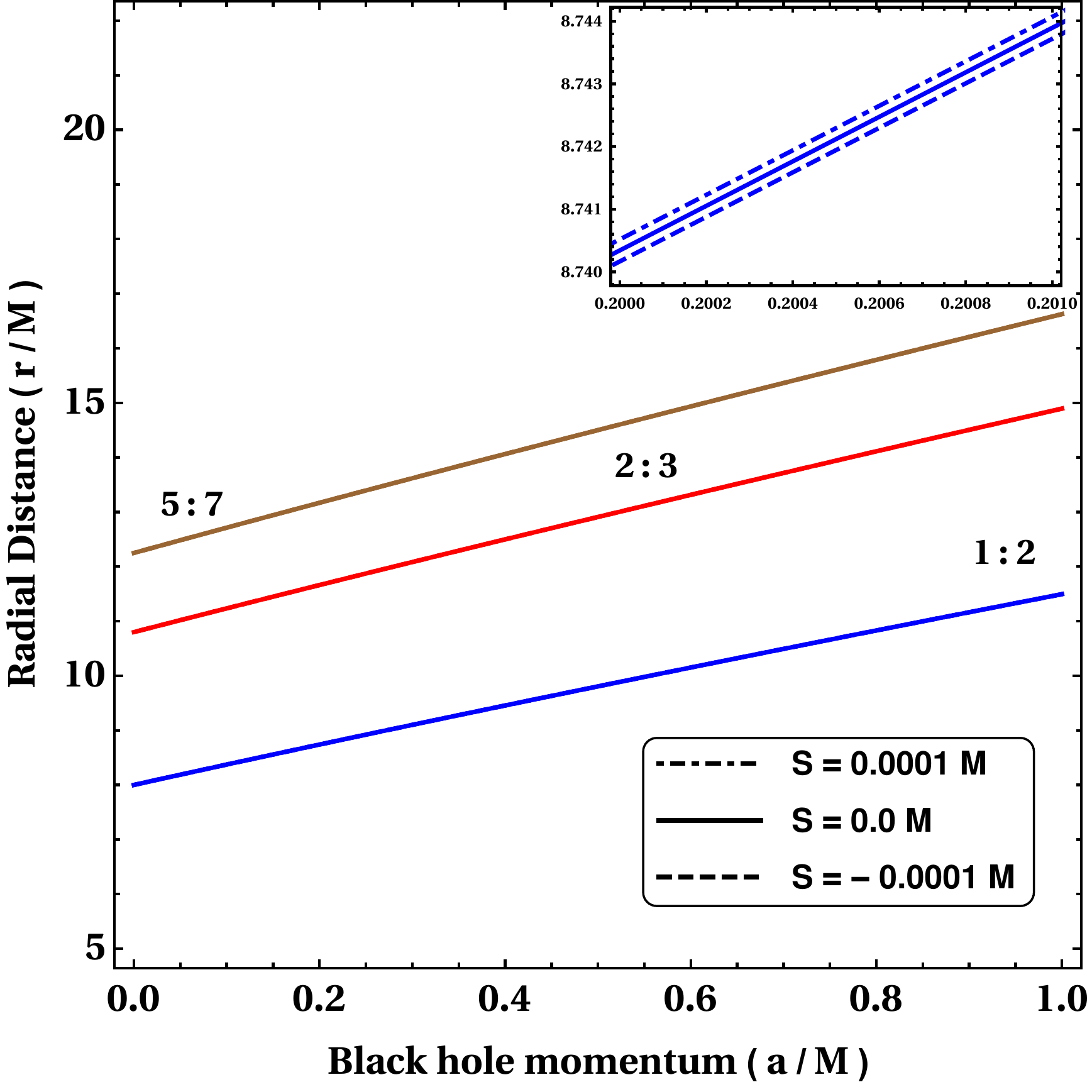}%
}\\
\centering
\subfloat[The above figure depicts the location of $\phi \theta$ resonant orbits for different spin of the secondary object.\label{fig:Resonance_02acnt}]{%
  \includegraphics[height=6.8cm,width=.46\linewidth]{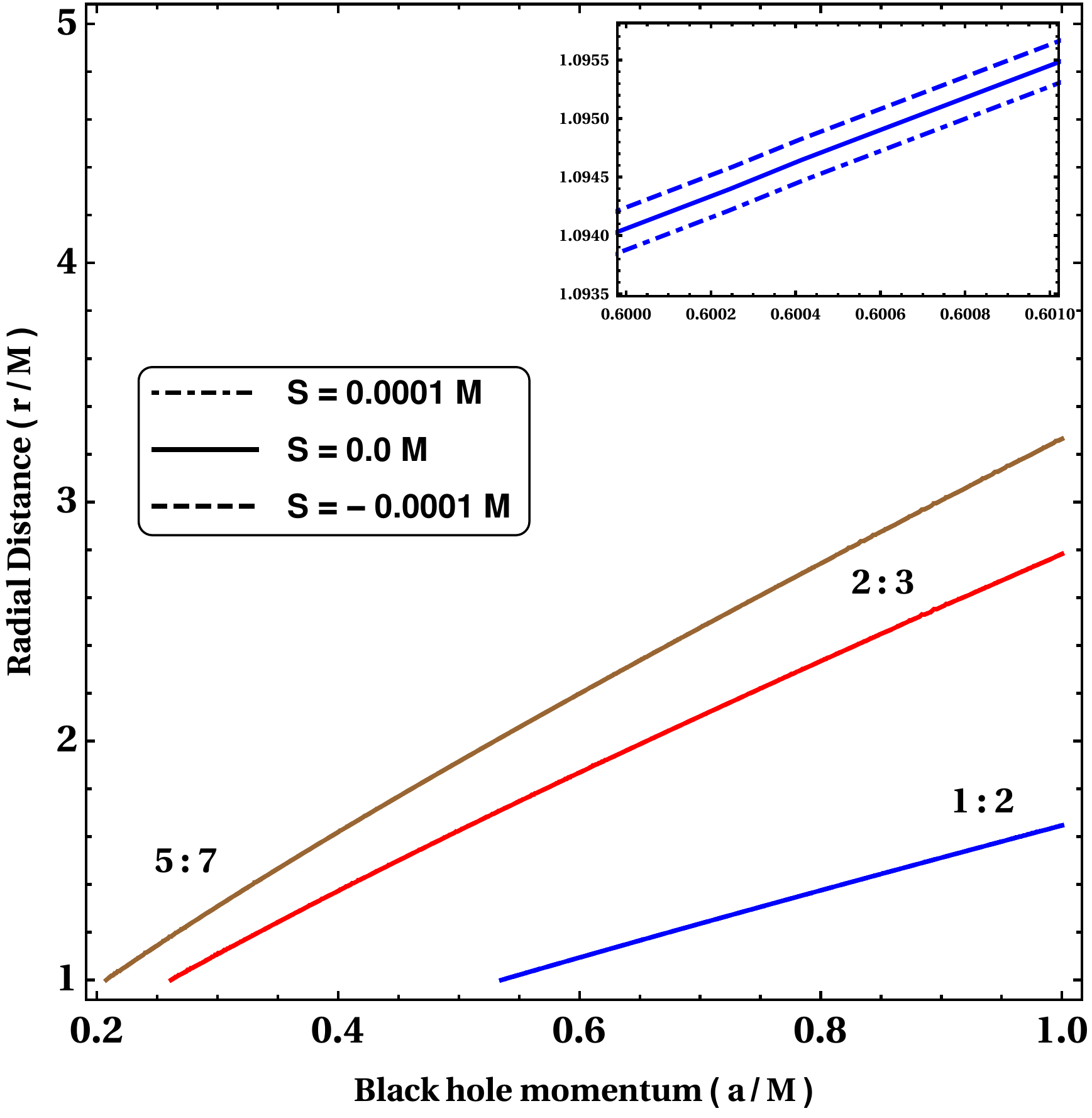}%
}\hfill
\caption{In the above figure, we have shown the counter-rotating resonant orbits as a function of the black hole's angular momentum, and different spin of the extended object. As expected, with the increase of the black hole's momentum, the orbits move away from the horizon.}
\label{fig:Resonance_Acnt}
\end{figure}

Having described various frequencies correspond to the small oscillation of a spinning particle, we can now demonstrate the resonant orbits depending on their spin parameters and black hole's angular momentum. In Fig. \ref{fig:Resonance_A}, the co-rotating resonant orbits are depicted as a function of the black hole's momentum while the spin parameter takes different values. The significance of the resonant orders are also demonstrated for $r\theta$, $r\phi$ and $\theta\phi$ resonances in Fig. \ref{fig:Resonance_01}, Fig. \ref{fig:Resonance_02}, Fig. \ref{fig:Resonance_02a} respectively. Given a larger order, the orbits move away from the horizon and as it is shown in Fig. \ref{fig:Resonance_A}, this is true for each of the resonance frequencies. Furthermore, it should also be emphasized that both $r\theta$ and $r\phi$ follow an identical  behavior while the $\theta\phi$ consists with a stark contrast from them. While in case of the $r\theta$ and $r\phi$ resonances, the orbits begin to move closer to the event horizon as one increases the black hole's momentum, nearly opposite phenomena happens for $\theta\phi$. This is related to the fact that a larger momentum of the black hole drags the prograde orbits close to its horizon while the opposite is true for retrograde orbits. Because of the presence of $r$ in both $r\theta$ and $r\phi$ resonances, this nature is largely influential in either of these cases. But in case of $\theta\phi$ resonance, this is no longer valid and  the orbits behaves quite differently from $r\theta$ and $\theta\phi$ resonances. Similarly, the resonant orbits for the counter-rotating trajectories are given in Fig. \ref{fig:Resonance_Acnt}. Interestingly, $\theta \phi$ resonance has now becomes $\phi\theta$! Mathematically, this is straightforward to understand from the relation given in Eq. (\ref{eq:fun_theta}), which stands as follows for the spinless case:
%%%%%%%%%%%%%%%%%
\begin{equation}
\dfrac{\Omega_{\phi}}{\Omega_{\theta}} =  \dfrac{r_{\rm c}}{\sqrt{3 a^2 \mp 4 a  \sqrt{M r_{\rm c}}+r^2_{\rm c}}}.
\end{equation}
%%%%%%%%%%%%%%%%% 
For the upper sign or co-rotating case, it is possible to find $\Omega_{\theta}<\Omega_{\phi}$, while in the case of counter-rotating case, $\Omega_{\theta}$ is always greater than $\Omega_{\phi}$. Therefore, in order to retain the convention used in Eq. (\ref{eq:primary}), we employ $\phi\theta$ resonance for the counter-rotating case. In the case of a spinning particle within the linear approximation, the above explanation remains the same.

With this, we finish our discussions regarding the resonance phenomenon between the small oscillations frequencies for a spinning particle with its spin approximated up to the linear order term. In the upcoming section, we consider the $r\phi$ resonance correspond to a generic trajectory confined on the equatorial plane of Kerr black hole.
%%%%%%%%%%%%%%%%%%%%%%%%%%%%
%%%%%%%%%%%%%%%%%%%%%%%%%%%%
%%%%%%%%%%%%%%%%%%%%%%%%%%%%
%%%%%%%%%%%%%%%%%%%%%%%%%%%%
%%%%%%%%%%%%%%%%%%%%%%%%%%%%%%%%%%%%%%%%%%%%%%%%%%%%
\section{Motion on the equatorial plane: the $r\phi$ resonance} \label{sec:rphi}
With the introduction to resonant orbits for spinning particle orbiting in 
nearly circular trajectories close to the equatorial plane, we grasp the basic 
mechanism to locate them as a function of the radial distance $r$. However, to 
address any general motion it is presumed that orbits are neither nearly 
circular nor close to the equatorial plane and an universal prescription is 
required to study them. In case of geodesics, these are well studied in 
literature since when Carter discovered the existence of a fourth conserved 
quantity and conclude that the geodesic trajectories are completely integrable 
in the Kerr background \cite{Carter:1969zz}. However, in the case 
of a spinning particle, general off-equatorial orbits can be nontrivial to obtain; recent studies on these aspects can be found in 
Refs. \cite{Witzany:2019dii,Witzany:2019nml}. Nonetheless for orbits confined 
on the equatorial plane, the \MP~equations can be exactly solved and thereby the 
resonance between the radial $\Omega_r$ and azimuthal $\Omega_{\phi}$ frequency 
can be well determined. The primary equation governing this 
resonance phenomena is given as $m \Omega_{r}-n\Omega_{\phi}=0$ and stated in 
Eq. (\ref{eq:primary}). From the solutions of \MP~equations, we can express 
$\dfrac{d\phi}{dr}$ as
%%%%%%%%%%%%%%%%%%%%%%%%%
\begin{equation}
\dfrac{d\phi}{dr}=\dfrac{\mathcal{U}^{3}}{\mathcal{U}^{1}}=\dfrac{1}{\sqrt{R_s}}\left[\left(1+\dfrac{3 M S^2}{r \Sigma_s}\right)\left\{J_z-(a+S)E\right\}+\dfrac{a}{\Delta}P_s\right],
\label{eq:dphidr}
\end{equation}
%%%%%%%%%%%%%%%%%%%%%%%%%
where $\mathcal{U}^{1}$ and $\mathcal{U}^{3}$ are radial and azimuthal velocity respectively and the quantities $P_s$, $R_s$, $\Sigma_s$ have been defined in Eq. (\ref{Eq:Ps}).
%%%%%%%%%%%%%%%%%%%%%%%%%

In the upcoming sections, we shall explicitly employ the above expression to locate the $r\phi$ resonance orbits in both Schwarzschild and Kerr geometries. However before studying any resonance phenomena, it is advisable to notice that the radial velocity $\mathcal{U}^{1}$ identically vanishes in the turning points $(r_a,r_p)$ and as a result the integral in Eq. (\ref{eq:primary}) would diverge. Therefore to compute the integral we may use an approximate technique and in the present context, Sochnev method can be extremely useful \cite{sochnev1968approximation} ( we also refer Ref. \cite{Kunst:2015hsa} where this is used in case of a spinning particle). For convenience, the basic design of Sochnev method is given below.

Let us consider an irrational function given as
%%%%%%%%%%%%%%%%%%%%%%%%%%
\begin{equation}
C=\sqrt{C_1 C_2....C_{\rm m}}
\end{equation}
%%%%%%%%%%%%%%%%%%%%%%%%%%  
and we desire to approximate it for our convenience. In the above, $C_1$, $C_2$,...$C_{\rm m}$ are positive numbers. The first step of Sochnev method \cite{sochnev1968approximation} dictates that the value of $C$ would fall within the upper limit $a_1$ and lower limit $b_1$ which are given by the following expressions
%%%%%%%%%%%%%%%%%%%%%%%%%%
\begin{equation}
a_1=\dfrac{C_1+C_2+.....C_m}{2}, \quad b_1=\dfrac{C_1C_2...C_m}{a_1^{m-1}}, \quad  \text{and} \quad b_1<C<a_1,
\end{equation}
%%%%%%%%%%%%%%%%%%%%%%%%%%
while more accuracy can be achieved with next order terms. The final expression of the sequence $(a_n,b_n)$ can be written as
%%%%%%%%%%%%%%%%%%%%%%%%%%
\begin{equation}
a_{n+1}=\dfrac{(m-1)a_n+b_n}{m}, \quad b_{n+1}=\dfrac{(a_n)^{m-1}b_n}{(a_{n+1})^{m-1}},
\end{equation}
%%%%%%%%%%%%%%%%%%%%%%%%%% 
and for large $n$, it is expected both $a_n$ and $b_n$ would generate extremely accurate values of $C$. However, in the present context we often neglect the higher order $n$ terms and truncate our series at $(a_1,b_1)$. Due to the nonzero spin terms the expressions become largely cumbersome to probe higher order terms with $n>1$.
%%%%%%%%%%%%%%%%%%%%%%%%%
%%%%%%%%%%%%%%%%%%%%%%%%%%%%%%%%%%%%%%%%%%%%%%%%%%%%%%%%%%%%%%%%%%%%%%%%%%%%%%%%%%%%%%%%%%%%
\subsection{In the Schwarzschild black hole}
Let us now describe the $r\phi$ resonance in the static and spherically symmetric Schwarzschild black hole and later on we discuss the Kerr geometry. For convenience, we start with the geodesic trajectories and then address the motion of a spinning particle.
%%%%%%%%%%%%%%%%%%%%%%%%%%%%%%%%%%%%%%%%%%%%%
\subsubsection{Geodesic limit:}
Due to the spherical symmetry, geodesic trajectories in \S~spacetime~are vastly simplified and easily obtainable in comparison to the rotating case. To study the resonance phenomena in the \S~background, we set $\theta=\pi/2$ without loosing any generality and obtain the radial potential as follows
%%%%%%%%%%%%%%%%%%%%%%%%%%
\begin{equation}
r^4 (\mathcal{U}^{1})^2=V(r)=E^2 r^4-r^2 L^2_z+2 M r L^2_z-(r^2-2 M r)r^2,
\label{eq:vr_sbh_geo}
\end{equation}
%%%%%%%%%%%%%%%%%%%%%%%%%%
where $E$ and $L_z$ are given as the conserved energy and momentum associated with the timelike and spacelike symmetries of the geometry respectively. Like mentioned earlier, at $r=r_a$ and $r=r_p$ the radial potential identically vanishes, i.e $V(r_a)=V(r_p)=0$, and we can arrive at the expressions for $E$ and $L_z$ in terms of $r_a$ and $r_p$. These are given by
%%%%%%%%%%%%%%%%%%%%%%%%%%
\begin{equation}
E^2_{\rm geo}=\dfrac{(r_a+r_p)(2M-r_a)(2M-r_p)}{r_a r_p (r_a+r_p)-2 M (r^2_a+r^2_p+r_a r_p)} \quad \text{and} \quad L^2_{\rm zgeo}=\dfrac{2 M r^2_a r^2_p}{r_a r_p (r_a+r_p)-2 M (r^2_a+r^2_p+r_a r_p)}.
\label{eq:conserver_geo_sbh}
\end{equation}
%%%%%%%%%%%%%%%%%%%%%%%%%%
Substituting these expression in the potential given by Eq. (\ref{eq:vr_sbh_geo}) we arrive at
%%%%%%%%%%%%%%%%%%%%%%%%%%
\begin{equation}
V(r)=-\dfrac{2M(r-r_a)(r-r_p)(r-0)(r-r_c)\left\{r_a r_p-2 M (r_a+r_p)\right\}}{r_a r_p (r_a+r_p)-2 M (r^2_a+r^2_p+r_a r_p)},
\label{eq:radial_potential_sbh}
\end{equation}
%%%%%%%%%%%%%%%%%%%%%%%%%%
where we have $r_c=\dfrac{2 M r_a r_p}{r_a r_p-2 M(r_a+r_p)}$ and the solutions follow $r_a > r_p > r_c$. From the expression, $ r_p > r_c$, we arrive at
%%%%%%%%%%%%%%%%%%
\begin{equation}
\dfrac{p-4M}{2M(1+e)}>1 \Longrightarrow p>2M(3+e).
\label{eq:pp}
\end{equation}
%%%%%%%%%%%%%%%%%%
Furthermore, with the expressions for square of momentum given in Eq. (\ref{eq:conserver_geo_sbh}), we require $L^2_{\rm zgeo}>0$. Therefore from the denominator, we gather
%%%%%%%%%%%%%%%%%%%%%%%%%%
\begin{equation}
r_a r_p (r_a+r_p)>2 M (r^2_a+r^2_p+r_a r_p),
\label{eq:constraint}
\end{equation}
%%%%%%%%%%%%%%%%%%%%%%%%%%
and from Eq. (\ref{eq:radial_potential_sbh}), we arrive at
%%%%%%%%%%%%%%%%%%%%%%%%%%
\begin{equation}
(r-r_a)(r-r_p)(r-r_c)\left\{r_a r_p-2 M (r_a+r_p)\right\}<0.
\label{eq:ra_rp}
\end{equation}
%%%%%%%%%%%%%%%%%%%%%%%%%%
By using Eq. (\ref{eq:pp}), one can establish $r_a r_p-2 M (r_a+r_p)>0$, and the above equation is automatically satisfied. Moreover, Eq. (\ref{eq:constraint}) can be further simplified by using, $r_{a}=p/(1-e)$ and $r_{p}=p/(1+e)$, and we arrive at
%%%%%%%%%%%%%%%%%%%%%%%%%%
\begin{equation}
\dfrac{2p^2 \bigl[p-M (3+e^2)\bigr]}{(1-e^2)^2}>0,
\end{equation}
%%%%%%%%%%%%%%%%%%%%%%%%%%
and for $p>M(3+e^2)$, it is always satisfied. Therefore, the final constraint on the semi-latus varies between $p=6M$ (for circular orbit) to $p=8M$ ( highest eccentric orbit $e=1$).
%%%%%%%%%%%%%%%%%%%%%%%%%%%%%%%%%%%%%%%%%%%%%
\subsubsection{Spinning particle:}

For a non-vanishing spin of the particle, the rescaled potential can be written as 
%%%%%%%%%%%%%%%%%%%%%%%%%%
\begin{equation}
V_{s}(r)=r^4 R_s=r^4 \biggl [(E^2-1)r^4+2 M r^3-r^2 J^2_{z}+2 r J_z \left\{M J_z+ES(r-3M)\right\} \biggr],
\end{equation}
%%%%%%%%%%%%%%%%%%%%%%%%%%
and by setting $S=0$, we get back the potential for a geodesic. Furthermore, it 
should be emphasized that the above equation contains terms only linear in $S$ 
while the higher order terms are ignored for a convenient computation. Naturally, one could ask whether the above potential gives the 
correct location for circular orbits or not. This could be easily verified as we 
already have the exact expression for radial potential for a spinning particle, 
given in Eq. (\ref{eq:eom}) and Eq. (\ref{Eq:Ps}). It can be shown that within the small spin approximation, i.e., $|S| \ll 1$, the above potential provides the locations 
of circular orbits within an infinitesimal mismatch from the exact derivation. 
For example, given a spin of $S=0.1M$ and energy $E=1$, the location of the 
circular orbits mismatch within $\mathcal{O}(10^{-3})$. With that spirit, we 
employ the above linearized expression and it simplifies our computations. We 
shall now concentrate in obtaining the expressions for energy and momentum 
assuming an ansatz of the form, $E_{\rm sbh}=E_{\rm geo}+S E_{\rm s}$ and 
$J_{\rm sbh}=L_{\rm zgeo}+S J_{\rm s}$ is valid. With the prior knowledge about 
both $E_{\rm geo}$ and $L_{\rm geo}$ \cite{wald1984general}, we can solve 
$V_{s}(r_a)=V_{s}(r_p)=0$ and obtain $E_{\rm s}$ and $J_{\rm s}$ in terms of 
$r_a$ and $r_p$. Both $E_{\rm }$ and $J_{\rm s}$ are given as follows 

%%

%%%%%%%%%%%%%%%%%%%%%%%%%%
\begin{eqnarray}
J_{\rm s} & = & \dfrac{\sqrt{(2 M - r_a) (2 M - r_p) (r_a + r_p)} \left[r_a r_p (r_a + r_p) - 
   3 M (r^2_a + r_a r_p + r^2_p)\right]}{\left[r_a r_p (r_a + r_p) - 
  2 M (r^2_a + r_a r_p + r^2_p)\right]^{3/2}}, \nonumber \\
  E_{\rm s} & = & \dfrac{-M\sqrt{2 M}r_a r_p}{\left[r_a r_p (r_a + r_p) - 
  2 M (r^2_a + r_a r_p + r^2_p)\right]^{3/2}},
\end{eqnarray}
%%%%%%%%%%%%%%%%%%%%%%%%%%
Further employing the above relations into the potential given upto the linear order in spin, we arrive at
%%%%%%%%%%%%%%%%%%%%%%%%%%
\begin{equation}
V_{\rm s}(r)=\dfrac{2 M r^5 a_0(r-r_a)(r-r_p)}{\left[r_a r_p (r_a+r_p)-2 M (r^2_a+r^2_p+r_a r_p)\right]^2}\left[ r+\dfrac{b_0}{a_0}\right],
\end{equation}
%%%%%%%%%%%%%%%%%%%%%%%%%%
where $a_0$ and $b_0$ are given by
%%%%%%%%%%%%%%%%%%%%%%%%%
\begin{eqnarray}
a_0 & = & -\left[r_a r_p-2 M (r_a+r_p)\right]\left\{r_a r_p(r_a+r_p)-2M(r^2_a+r^2_p+r_a r_p)\right\} \nonumber \\
 {}{} & & -S r_a r_p\left\{2M(r_a-2M)(r_p-2M)(r_a+r_p)\right\}^{1/2},\nonumber \\
b_0 & = & 2 M r_a r_p \left\{r_a r_p(r_a+r_p)-2M(r^2_a+r^2_p+r_a r_p)\right\}\nonumber \\
& & -S r_a r_p (r_a+r_p)\left\{2M(r_a-2M)(r_p-2M)(r_a+r_p)\right\}^{1/2}.\nonumber \\
\end{eqnarray}
%%%%%%%%%%%%%%%%%%%%%%%%%
It should be noted that even if the number of solutions remain identical with the geodesic case, there now exists explicit spin dependence in one of the solutions $r_s=-b_0(a_0)^{-1}$. Finally, the potential can be compactly written in the form of
%%%%%%%%%%%%%%%%%%%%%%%%
\begin{equation}
V_{r}(s)=\dfrac{2 M a_0r^4(r-0)(r-r_a)(r-r_p)(r-r_s)}{\left[r_a r_p (r_a+r_p)-2 M (r^2_a+r^2_p+r_a r_p)\right]^2},
\end{equation}
%%%%%%%%%%%%%%%%%%%%%%%%
and identical to the geodesic case, we require to have $V_{r}(s)>0$. This condition along with the bound $r_p<r<r_a$ would give $r_p>r_s$ and $a_0<0$. However either of these constraints would depend on the numerical value of the spin parameter and therefore while obtaining the resonant orbits, we will employ it explicitly.  

With the above mentioned points kept in mind, we now determine $\mathcal{U}^{3}(\mathcal{U}^1)^{-1}$ from Eq. (\ref{eq:dphidr}) and in the linear spin approximation, this is given as
%%%%%%%%%%%%%%%%%%%%%%%%
\begin{equation}
\dfrac{d\phi}{dr}=\dfrac{(J_z-ES)\left\{r_a r_p (r_a+r_p)-2 M (r^2_a+r^2_p+r_a r_p)\right\}}{\biggl[2 M a_0(r-0)(r-r_a)(r-r_p)(r-r_s)\biggr]^{1/2}}.
\end{equation}
%%%%%%%%%%%%%%%%%%%%%%%%
By substituting it in Eq. (\ref{eq:primary}), we arrive at the following expression which describes the $r\phi$ resonance for a given particle
%%%%%%%%%%%%%%%%%%%%%%%%
\begin{equation}
m\pi-n \int^{r_a}_{r_p}\biggl(\dfrac{d\phi}{dr}\biggr)dr=m\pi-n I_s = 0.
\label{eq:resonance_condition}
\end{equation}
%%%%%%%%%%%%%%%%%%%%%%%%
In the discussions given below, we employ the \SV~formalism that is earlier introduced in section-(\ref{sec:rphi}) and compute $I_s$ approximately. However for a convenient and simplified computation, we confine our discussions only with the first order approximation, i.e, we assume $(C_1 C_2)^{1/2} \approx 2^{-1}(C_1+C_2)$ is valid throughout our calculations.

\begin{figure}[htp]
\subfloat[The orbits for 2:3 resonance is given for both spinning and non-spinning particle. For a negative spin parameter of the particle, the resonant orbits appear at a larger range of $p$ while for positive spin, the opposite phenomenon appears.\label{fig:Resonance_03}]{%
  \includegraphics[height=6cm,width=.49\linewidth]{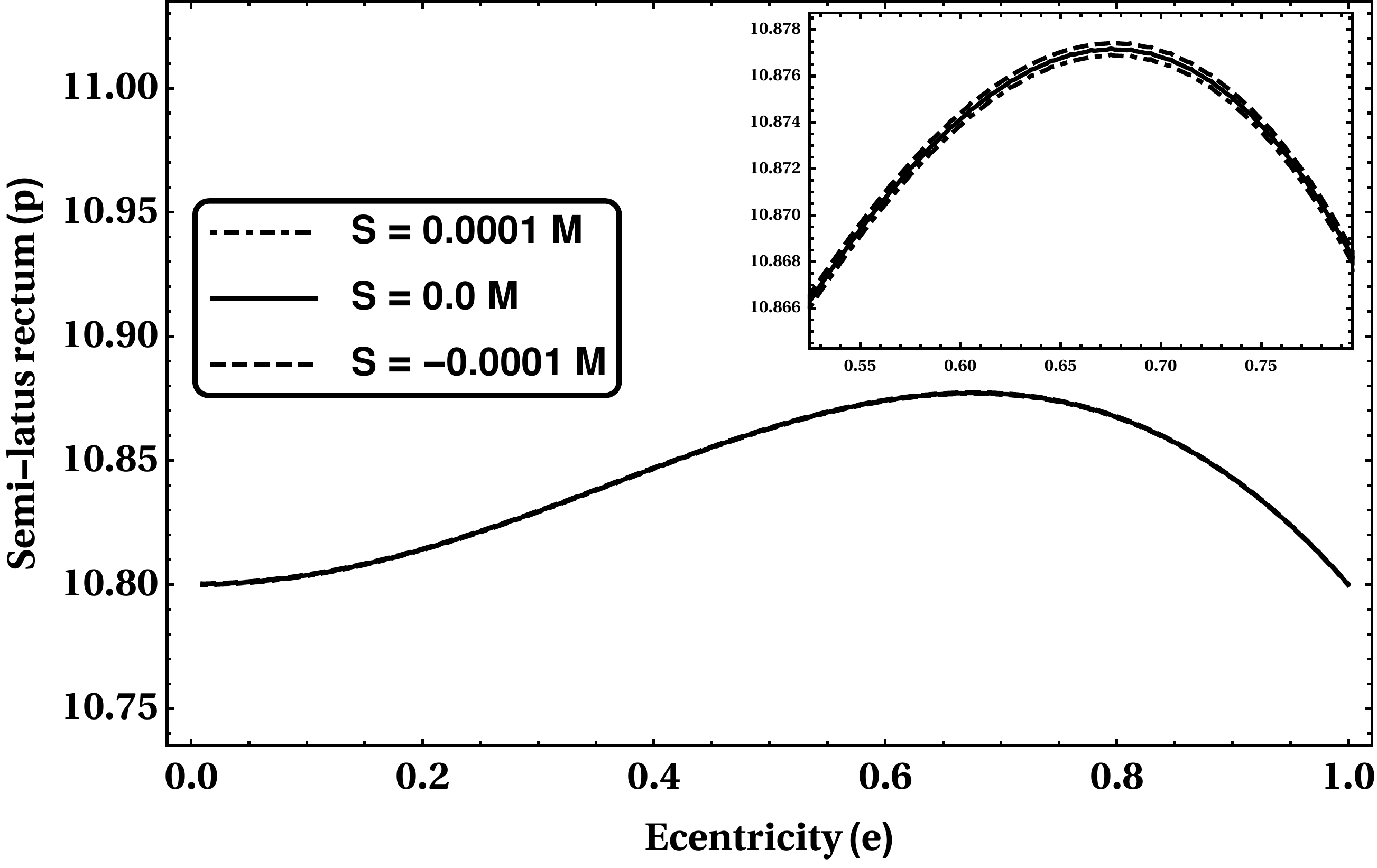}%
}\hfill
\subfloat[Orbits for 1:2 resonance are given in a \S~background. The disntinguishing features for both positive and negative values of spin are given explicitly. \label{fig:Resonance_04}]{%
  \includegraphics[height=6cm,width=.49\linewidth]{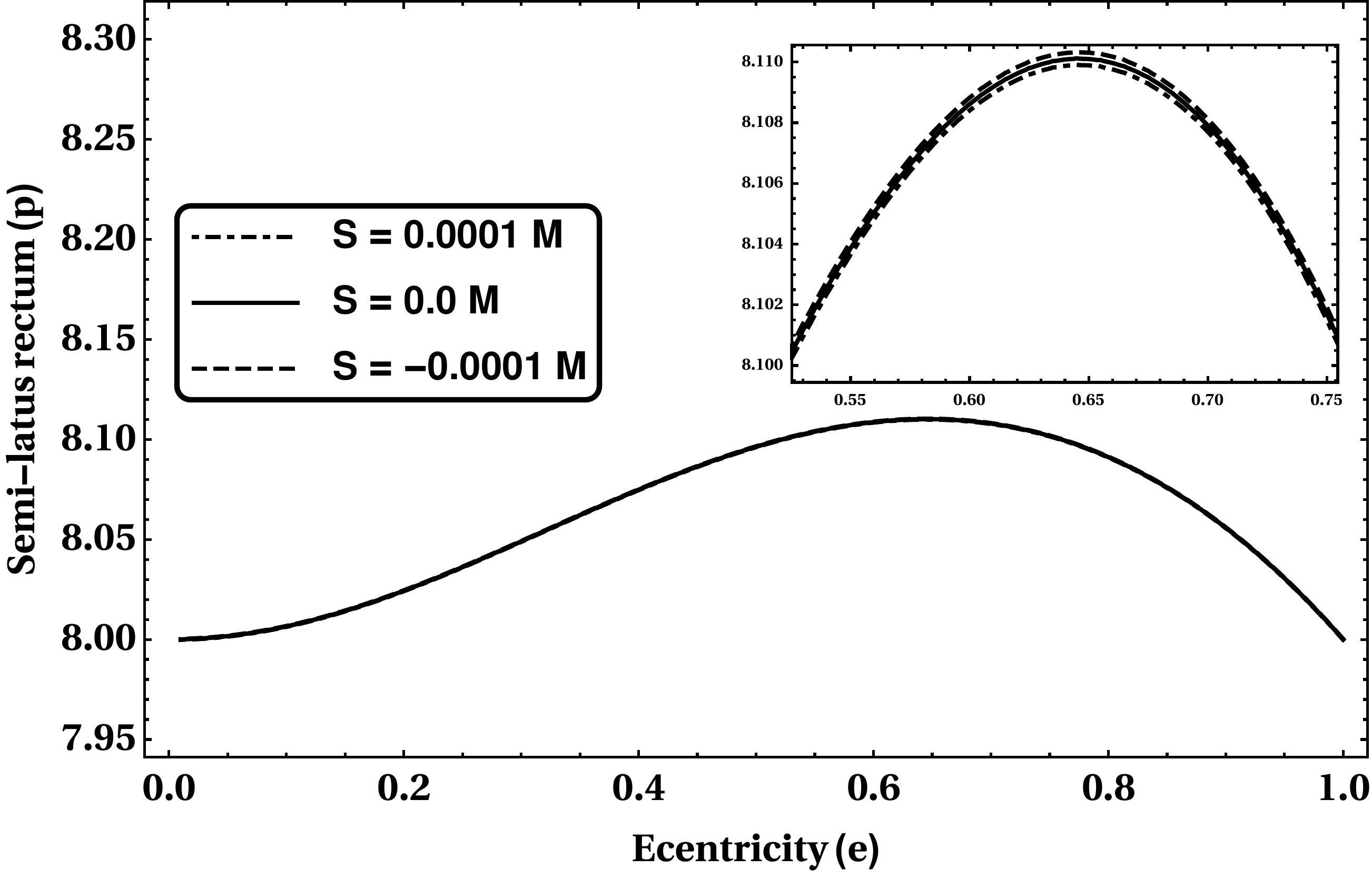}%
}
\caption{The above figure demonstrates various $r\phi$ resonant orbits in a \S~spacetime for both spinning and non-spinning trajectories. In the inset, we zoom a particular segment to show the difference for various spin parameters.}
\label{fig:Resonance_B}
\end{figure}

In order to obtain an approximate value for the above integral, we start with
%%%%%%%%%%%%%%%%%%%%%%%
\begin{equation}
\int^{r_a}_{r_p}\dfrac{d\phi}{dr}dr=N \int^{r_a}_{r_p}\dfrac{dr}{\sqrt{a_0(r-0)(r-r_a)(r-r_p)(r-r_s)}},
\end{equation}
%%%%%%%%%%%%%%%%%%%%%%%
where $N=(2M)^{-1/2}(J_z-ES)\left\{r_a r_p (r_a+r_p)-2M(r^2_a+r^2_p+r_a r_p)\right\}$ and $a_0$ is already defined earlier. Let us now introduce a coordinate transformation given as $r=2^{-1}\bigl[(r_a+r_p)+x(r_a-r_p)\bigr]$ and with that, we arrive at
%%%%%%%%%%%%%%%%%%%%%%%
\begin{equation}
I_s=\int^{r_a}_{r_p}\left(\dfrac{d\phi}{dr}\right)dr=\int^{1}_{-1}\left(\dfrac{d\phi}{dx}\right)dx.
\end{equation}
%%%%%%%%%%%%%%%%%%%%%%% 
Finally we have 
%%%%%%%%%%%%%%%%%%%%%%% 
\begin{equation}
\dfrac{d\phi}{dx}=\dfrac{r_a-r_p}{2}\dfrac{d\phi}{dr}=\dfrac{r_a-r_p}{2}\dfrac{N}{\left\{-a_0 \dfrac{(r_a-r_p)^2}{4}(1-x^2)(r-0)(r-r_s)\right\}^{1/2}},
\end{equation}
%%%%%%%%%%%%%%%%%%%%%%% 
and after further simplification, we write it as follows
%%%%%%%%%%%%%%%%%%%%%%% 
\begin{eqnarray}
\dfrac{d\phi}{dx} &=& \dfrac{N}{\left\{-a_0(1-x^2)(r-0)(r-r_s)\right\}^{1/2}}, \nonumber \\
&=& {N}{\left\{-a_0\dfrac{(r_a+r_p)}{2}\dfrac{(r_a+r_p-2r_s)}{2}(1-x^2)(1+K_0 x)(1+K_s x)\right\}^{-1/2}}.\\
\end{eqnarray}
%%%%%%%%%%%%%%%%%%%%%%% 
With $<N>$ given by
%%%%%%%%%%%%%%%%%%%%%%%%
\begin{equation}
<N>={2N}{\left\{-a_0 (r_a+r_p)(r_a+r_p-2r_s)\right\}^{-1/2}}, 
\end{equation}
%%%%%%%%%%%%%%%%%%%%%%%%
the above can be written in a more compact form as
%%%%%%%%%%%%%%%%%%%%%%%
\begin{equation}
\dfrac{d\phi}{dx}=\dfrac{<N>}{\left\{(1-x^2)(1+K_0 x)(1+K_s x)\right\}^{1/2}},
\label{eq:dphix}
\end{equation}
%%%%%%%%%%%%%%%%%%%%%%%
where we have, $K_0=\dfrac{r_a-r_p}{r_a+r_p}$ and $K_{s}=\dfrac{r_a-r_p}{r_a+r_p-2r_s}$. Furthermore, to ease our computations, we introduce the expression
%%%%%%%%%%%%%%%%%%%%%%%
\begin{equation}
\int^{1}_{-1}\dfrac{1}{\sqrt{1-x^2}(1+K x)}dx=\dfrac{\pi}{\sqrt{1-K^2}},
\end{equation}
%%%%%%%%%%%%%%%%%%%%%%%
which is only valid with $K<1$. We can use the same technique in our case by writing $\sqrt{(1+K_0 x)(1+K_s x)} \approx \biggl(1+\dfrac{K_0+K_s}{2}\biggr)=1+K x$ and therefore, the final expression becomes
%%%%%%%%%%%%%%%%%%%%%%%
\begin{equation}
I_s=\int^{1}_{-1}\dfrac{d\phi}{dx}dx=\dfrac{\pi <N>}{\sqrt{1-K^2}}.
\end{equation}
%%%%%%%%%%%%%%%%%%%%%%%
With the above relation and by using Eq. (\ref{eq:resonance_condition}), the resonance condition takes the form 
%%%%%%%%%%%%%%%%%%%%%%%
\begin{equation}
m-n\dfrac{<N>}{\sqrt{1-K^2}}=0.
\end{equation}
%%%%%%%%%%%%%%%%%%%%%%%
We shall now employ the above condition and locate the resonant orbits for a spinning particle moving around a \S~black hole. In Fig. \ref{fig:Resonance_B}, we demonstrate the same for both 2:3 (Fig. \ref{fig:Resonance_03}) as well as 1:2 (Fig. \ref{fig:Resonance_04}) resonance condition.
%%%%%%%%%%%%%%%%%%%%%%%%%%%%
%%%%%%%%%%%%%%%%%%%%%%%%%%%%
%%%%%%%%%%%%%%%%%%%%%%%%%%%%%%%%%%%%%%%%%%%%%%%%%%%%%%%%%%%%%%%%%%%%%%%%%%%%%%%%%%%%%%%%%%%%

%%%%%%%%%%%%%%%%%%%%%%%%%%%%%%%%%%%%%%%%%%%%%%%%%%%%%%%%%%%%%%%%%%%%%%%%%%%%%%%%%%%%%%%%%%%%
\subsection{In the Kerr black hole}
Having described the resonance phenomena in \S~background, we shall continue our discussions in the Kerr spacetime and as expected, in rotating geometry the motion of a spinning particle is additionally complicated. Therefore before delving into the details of spinning particle, we start with the geodesic case and then address the extended object in both linearized Kerr and full Kerr spacetime. 

%We start by recapitulating the geodesic case and then study the linearized Kerr spacetime for a spinning particle and then finally, we shall address the resonance phenomena for an extended object orbiting in a Kerr spacetime. 
%%%%%%%%%%%%%%%%%%%%%%%%%%%%%%%%%%%%%%%%%%%%%%%%%%%%%%%%%%%%%%%%%%%%%%%%%%%%%%%%%%%%%%%%%%%%
\subsubsection{Geodesic limit:}
\begin{figure}[htp]
\subfloat[The resonant orbits are given for $a=0.1M$. With the increase in ecentricity, the semi-latus rectum increases and reaches a maximum value at $p=p^{\rm max}$ while further increase in ecentricity decreases $p$ as well \label{fig:Resonance_05}]{%
  \includegraphics[height=6.2cm,width=.49\linewidth]{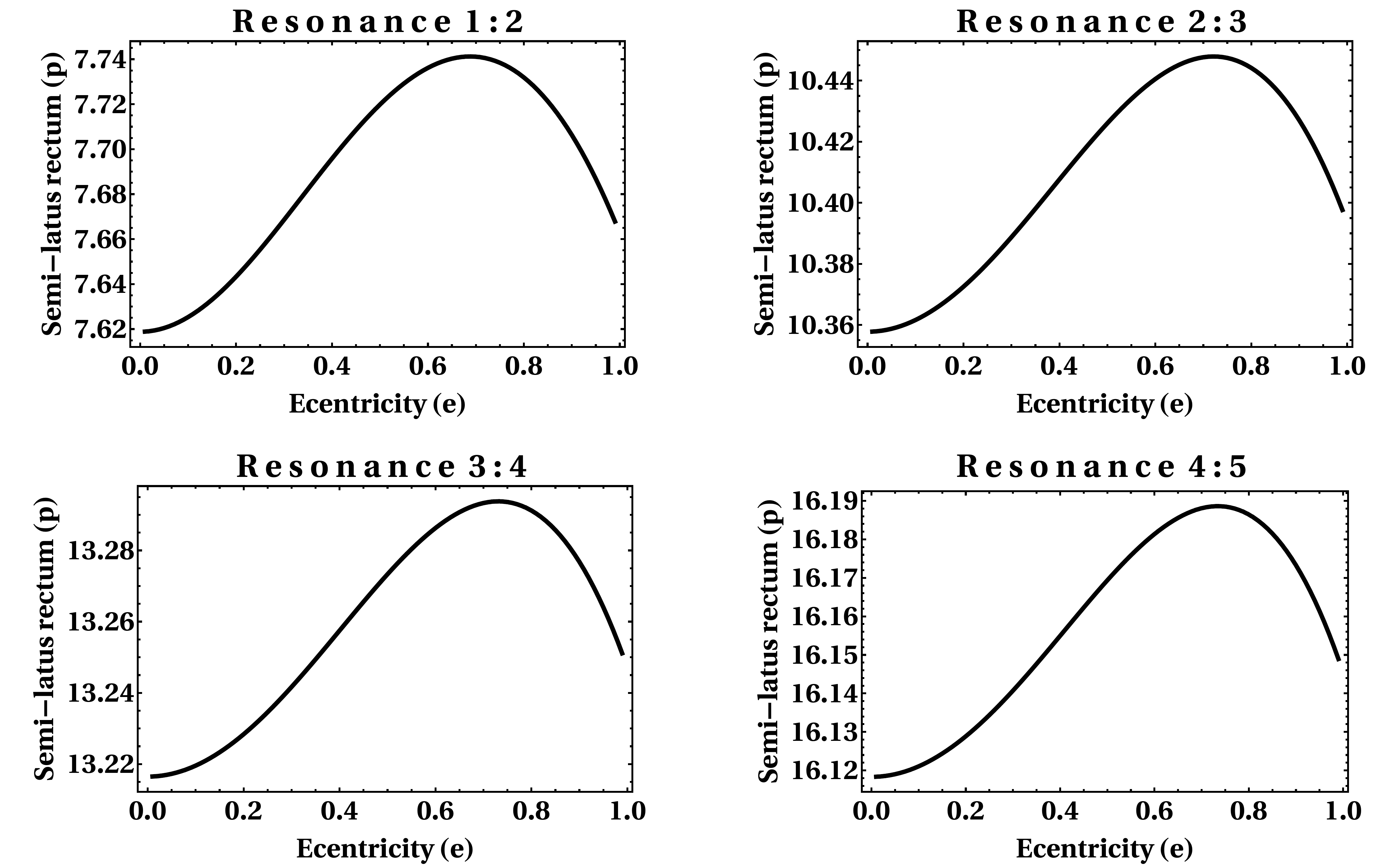}%
}\hfill
\subfloat[Figure shows the resonant orbits for a geodesic while the background is given by a Kerr geometry with $a=0.9M$. The nature of the plot changes drastically from Fig. \ref{fig:Resonance_05}. \label{fig:Resonance_06}]{%
  \includegraphics[height=6.2cm,width=.49\linewidth]{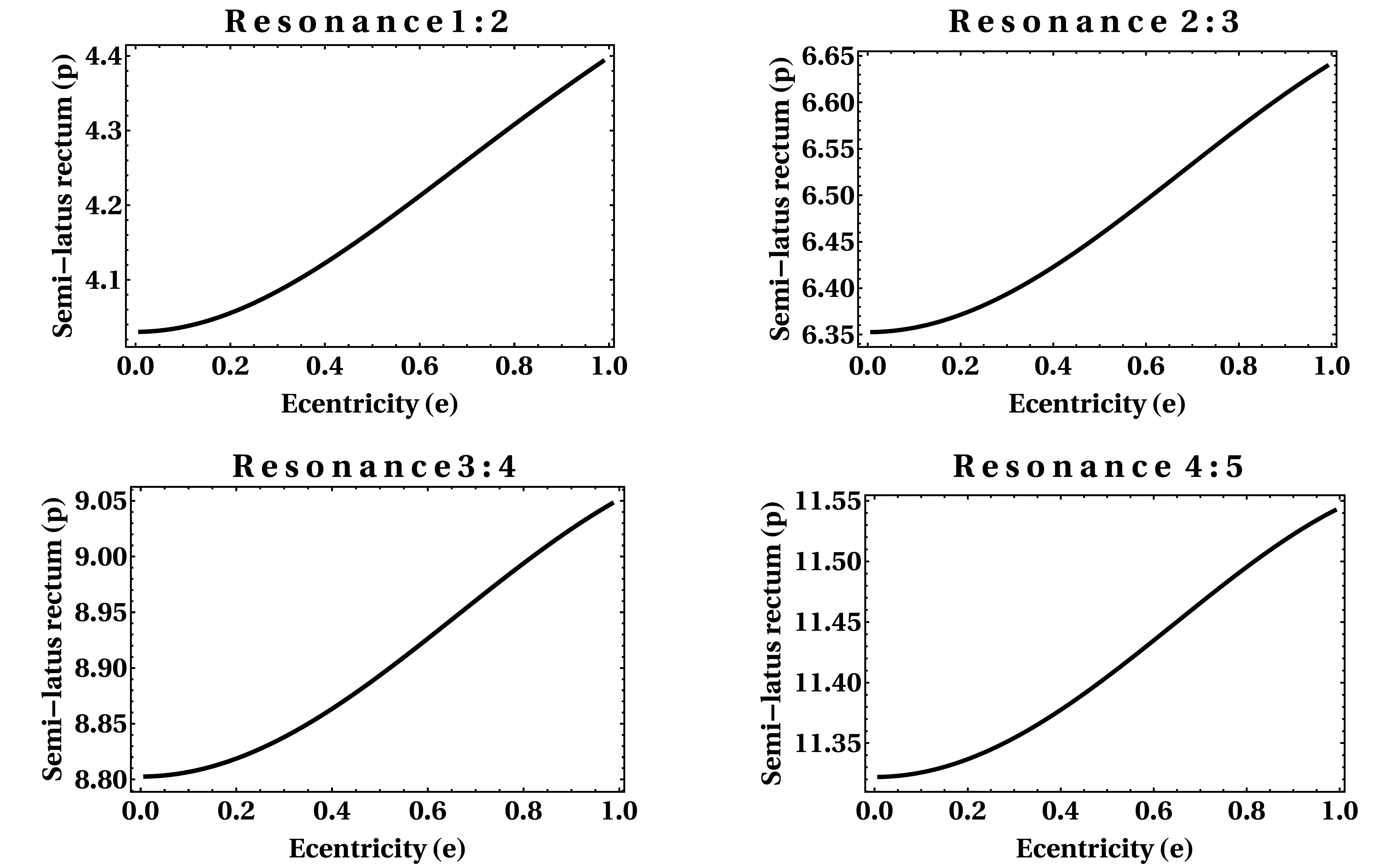}%
}
\caption{The above figure demonstrates various $r\phi$ resonant orbits in the Kerr spacetime for non-spinning trajectories.}
\label{fig:Resonance_C}
\end{figure}

To locate the resonant orbits for a geodesic, we need to employ the following relation
%%%%%%%%%%%%%%%%%%%%%%%%%
\begin{equation}
\dfrac{d\phi}{dr}=\dfrac{\mathcal{U}^{3}}{\mathcal{U}^{1}}=\dfrac{r^2 \left\{L_z-aE+\dfrac{a}{\Delta}\biggl[E(r^2+a^2)-a L_z\biggr]\right\}}{\biggl\{r^4 \biggl[E(r^2+a^2)-aL_z \biggr]^2-\Delta \biggl[r^6+(L_z-aE)^2r^4\biggr]\biggr\}^{1/2}},
\end{equation}
%%%%%%%%%%%%%%%%%%%%%%%%%
in Eq. (\ref{eq:primary}). With a more simplified form, the above can be written as
%%%%%%%%%%%%%%%%%%%%%%%%%
\begin{equation}
\dfrac{d\phi}{dr}=\dfrac{\left\{L_z-aE+\dfrac{a}{\Delta}\biggl[E(r^2+a^2)-a L_z\biggr]\right\}}{\biggl\{ \biggl[E(r^2+a^2)-aL_z \biggr]^2-\Delta \biggl[r^2+(L_z-aE)^2\biggr]\biggr\}^{1/2}}=\dfrac{\left\{L_z-aE+\dfrac{a}{\Delta}\biggl[E(r^2+a^2)-a L_z\biggr]\right\}}{\biggl\{ V(r)\biggr\}^{1/2}},
\end{equation}
%%%%%%%%%%%%%%%%%%%%%%%%%
and the potential takes the form,
%%%%%%%%%%%%%%%%%%%%%%%%%
\begin{equation}
V(r)=\biggl[E(r^2+a^2)-aL_z \biggr]^2-\Delta \biggl[r^2+(L_z-aE)^2\biggr].
\end{equation}
%%%%%%%%%%%%%%%%%%%%%%%%%
By further simplification, we arrive at 
%%%%%%%%%%%%%%%%%%%%%%%%%
\begin{equation}
V(r)=r\left[E^2 r^3-(L^2_z-a^2E^2)r+2 M (L_z-aE)^2-(r^2-2Mr+a^2)r\right]=-\alpha (r-0)(r-r_a)(r-r_p)(r-r_1), \\ 
\\
\end{equation}
%%%%%%%%%%%%%%%%%%%%%%%%%
with $\alpha=1-E^2$. From the above equation, we can now match the coefficients 
of $r^3$, $r^2$ and $r$ from each side. These would reproduce three independent 
equations and we can solve for energy, momentum and $r_1$ in terms of $r_a$ and 
$r_p$. Finally, we have 
%%%%%%%%%%%%%%%%%%%%%%%%
\begin{equation}
E^2=1-\dfrac{2M}{r_a+r_p+r_1}, \quad \text{and} \quad L^2_z=\dfrac{2M}{r_a+r_p+r_1} \left\{r_a r_p+r_1(r_a+r_p)-a^2\right\},
\end{equation}
%%%%%%%%%%%%%%%%%%%%%%%%
and the solution $r_1$ can be evaluated from the other condition
%%%%%%%%%%%%%%%%%%%%%%%%
\begin{equation}
4 a M E L_z=2 M a^2 E^2+2 M L^2_z-(1-E^2)r_1 r_a r_p.
\label{eq:ELZ}
\end{equation}
%%%%%%%%%%%%%%%%%%%%%%%%
By solving the above equation numerically for a given eccentricity $e$ and semi-latus rectum $p$, we obtain $r_1$ and therefore, determine both energy and momentum corresponding to the orbit. Henceforth, for any provided $r_a$ and $r_p$ we can obtain the conserved quantities require to specify the geodesic trajectory. It should be reminded that Eq. (\ref{eq:ELZ}) is a quadratic of $r_1$ and these two solutions represent either prograde or retrograde orbits. Like mentioned earlier, as we are interested in the prograde orbit, we would suppress the retrograde part and continue our analysis with that motivation. Nonetheless, a similar study can be carried in the case of retrograde orbits too. The next task is to compute the integral given in Eq. (\ref{eq:primary}) and locate the resonant orbits for an arbitrary order and this is carried out in the Appendix for both spinning and non-spinning trajectories. In Fig. \ref{fig:Resonance_C}, we have shown the resonant orbits for a geodesic moving in a spacetime with different rotation parameters $a$. For smaller values of $a$, the resonant orbits lie away from the black hole horizon and the nature remains similar to the \S~case. With the increase of resonance order, i.e, $m+n$, the orbits keep moving away from the horizon but contained with identical properties as given in Fig. \ref{fig:Resonance_05}. However for larger momentum, the peak seems to disappears and the Gaussian nature changes to linear curve. This is explicitly shown in Fig. \ref{fig:Resonance_06}.

%Apparently it may look like that the with larger eccentricity, the orbits get closer to the horizon of the black hole as $p$ decreases. However, this is not the case as there exists $1-e$ in the denominator and for values of $e$ close to unity, $r_a$ would be large. Therefore, more eccentric orbits would be assigned with larger $R=r_a-r_p$ values and so on. 
%It is schematically shown in Fig-()

%%%%%%%%%%%%%%%%%%%%%%%%%%%%
%%%%%%%%%%%%%%%%%%%%%%%%%%%%
%%%%%%%%%%%%%%%%%%%%%%%%%%%%%%%%%%%%%%%%%%%%%%%%%%%%%%%%%%%%%%%%%%%%%%%%%%%%%
\subsubsection{Spinning particle: In the slowly rotating Kerr black hole}
%%%%%%%%%%%%%%%%%%%%%%%
With the discussions presented above for geodesic trajectories, we shall now consider the case with spinning particles and locate their resonance orbits. However before dealing with the Kerr spacetime consists with the multipole structure of arbitrary order, we first consider only the dipole current moment and approach analytically to obtain the expressions for energy and momentum. In the usual Boyer-Lindquist coordinates, the resultant metric takes the following form
%%%%%%%%%%%%%%%%%%%%%%
\begin{equation}
ds^2=-\Bigl(1-\dfrac{2 M }{r} \Bigr)dt^2+{\Bigl(1-\dfrac{2 M }{r}\Bigr)}^{-1}{dr^2}+r^2 (d\theta^2+\sin^2\theta d\phi^2)-\dfrac{4 M a \sin^2\theta}{r}d\phi dt,
\end{equation}
%%%%%%%%%%%%%%%%%%%%%% 
with $a$ and $M$ given as the angular momentum per mass and mass of the spacetime respectively. For the above metric, we obtain the radial potential written up to the terms linear in $S$ as
%%%%%%%%%%%%%%%%%%%%%%%
\begin{eqnarray}
V_{\rm s}(r)=r^3 \biggl[r \left\{E^2 r^4-J_z^2 r^2+2 M r J_z^2-4 a M r E J_z-(r^2-2 M r)r^2\right\}+ \nonumber \\
S (2 E J_z r^3-6 E M J_z r^2+2 a M J_z^2+6 M a E^2 r^2)\biggr],
\end{eqnarray}
%%%%%%%%%%%%%%%%%%%%%%% 
and after further simplifications, we arrive at the following expression
%%%%%%%%%%%%%%%%%%%%%%%
\begin{eqnarray}
V_{\rm s}(r)&=&r^3 \biggl[r \left\{(E^2-1)r^4+2 M r^3-r^2 J_z^2+2 r \biggl(M J_z^2+EJ_z S (r-3 M)+3 a M E^2 S-2 a M E J_z\biggr)\right\} \nonumber \\ 
 &&  \hspace{10cm}+ 2 a M S J_z^2\biggr].
\label{eq:potential_Kerr}
\end{eqnarray}
%%%%%%%%%%%%%%%%%%%%%%%
We shall now assume, $E=E_{\rm sbh}+a E_{\rm s}$ and $J_z=J_{\rm sbh}+a J_{\rm 
s}$ and solve for $E_{\rm s}$ and $J_{\rm s}$ from the equations $V_{\rm 
s}(r_a)=V_{\rm s}(r_p)=0$. Such assumptions would ensure that 
only the terms linear to $a$ are considered, while higher 
order terms, 
i.e., $\mathcal{O}(a^2)$ and beyond, are neglected. In order to deal with 
larger momentum of the black hole, we carry out a numerical analysis in 
the 
next section. Due to the complexity of the forms of both $E_{\rm s}$ and 
$J_{\rm s}$, they are no explicitly written in the present paper. By 
substituting them in Eq. (\ref{eq:potential_Kerr}), we can write down the 
potential $V_{\rm s}(r)$ as
%%%%%%%%%%%%%%%%%%%%%%%
\begin{equation}
V_{\rm s}(r)=r^3\biggl[V^{\rm sbh}_{\rm s}(r)+a V^{\rm kerr}_{\rm s}(r)\biggr],
\end{equation}
%%%%%%%%%%%%%%%%%%%%%%%
where, $V^{\rm sbh}_{\rm s}(r)$ and $V^{\rm kerr}_{\rm s}(r)$ can be written as
%%%%%%%%%%%%%%%%%%%%%%%
\begin{eqnarray}
V^{\rm sbh}_{\rm s}(r) &= & \dfrac{2 M r^2 (r-r_a)(r-r_p)}{\mathcal{X}^{3}}\biggl[\mathcal{X}^{2}\biggl(2 M \{r_a r_p +r(r_a+r_p)\}-r r_a r_p\biggr)-r_a r_p (r+r_a+r_p)\mathcal{Y}^{1/2}\mathcal{X}S\biggr], \nonumber \\
V^{\rm kerr}_{\rm s}(r) &= & \dfrac{2M(r-r_a)(r-r_p)F(r)}{\mathcal{X}^3},
\end{eqnarray}
%%%%%%%%%%%%%%%%%%%%%%%
and $F(r)$, $\mathcal{X}$ and $\mathcal{Y}$ has the following expressions
%%%%%%%%%%%%%%%%%%%%%%%
\begin{eqnarray}
F(r) &= & -2 r^2 r_a r_p \mathcal{X}\mathcal{Y}(r+r_a+r_p)+r^2 \biggl[\biggl(2M \{r_a r_p +r (r_a+r_p)\}-r r_a r_p\biggr)\mathcal{X}^2-\nonumber \\
& & \hspace{10cm} r_a r_p (r+r_a +r_p)\mathcal{X}\mathcal{Y}S\bigr], \nonumber \\
\mathcal{X} &= &r_a r_p (r_a+r_p)-2M(r_a^2+r_a r_p+r_p^2), \qquad  \mathcal{Y}={2M(2M-r_a)(2M-r_p)(r_a+r_p)}. \nonumber \\
\end{eqnarray}
%%%%%%%%%%%%%%%%%%%%%%%%
Therefore, the complete form of the potential can be written as
%%%%%%%%%%%%%%%%%%%%%%%%
\begin{eqnarray}
V_{\rm s}(r) & = & \dfrac{2Mr^3(r-r_a)(r-r_p)}{\mathcal{X}^3}\biggl[F(r)+r^2\mathcal{X}^{2}\biggl(2 M \{ r_a r_p +r(r_a+r_p)\}\biggr)-r^2r_a r_p (r+r_a+r_p)\mathcal{Y}^{1/2}\mathcal{X}S\biggr],\nonumber \\
%&=&\dfrac{2Mr^3(r-r_a)(r-r_p)}{\mathcal{X}^3}\biggl[\alpha^{\prime} r^3+\beta r^2+\gamma r+\delta\biggr], \nonumber \\
&=& \dfrac{2Mr^3(r-r_a)(r-r_p)}{\mathcal{X}^3}\biggl[\alpha^{\prime} (r-r_1)(r-r_2)(r-r_3)\biggr], \nonumber \\
&=& {-\alpha r^3}(r-r_a)(r-r_p)(r-r_1)(r-r_2)(r-r_3), \nonumber \\
%&=& \dfrac{2M(r-r_a)(r-r_p)}{\mathcal{X}^3}\biggl[\alpha (r-r_3)(r^2-r r_{+}+r_{\times})\biggr].
\end{eqnarray}
%%%%%%%%%%%%%%%%%%%%%%%%
%where, $r_{+}=r_1+r_2$ and $r_{\times}=r_1 r_2$. In principle, one can solve for $r_{+}$ and $r_{\times}$ and write them in terms of $\alpha$, $\beta$, $\gamma$ and $\delta$.
where, $\alpha=-2 M\alpha^{\prime}/\mathcal{X}^3$. In principle, one can solve for $r_1$, $r_2$ and $r_3$ for different values of $r_a$ and $r_p$. This way, we can obtain the resonant orbits for a spinning particle while neglecting the higher spin contribution, i.e $\mathcal{O}(a^2)$, from the black hole. In the next discussion, we consider the Kerr black hole and numerically present the resonant orbits for a particle with spin.
%%%%%%%%%%%%%%%%%%%%%%%%%%%%%%%%%%%%%%%%%%%%%%%%%%%%%%%%%%%%%%%%%%%%%%%%%%%%%%%%%%%%%%%%%%%%
\subsubsection{Spinning particle: the Kerr spacetime}\label{sec:spin_arbitrary_spin}
\begin{figure}[htp]
\subfloat[The resonant orbits for a spinning particle is shown in a black hole with angular momentum $a=0.099M$. \label{fig:Resonance_07}]{%
  \includegraphics[height=6.2cm,width=.48\linewidth]{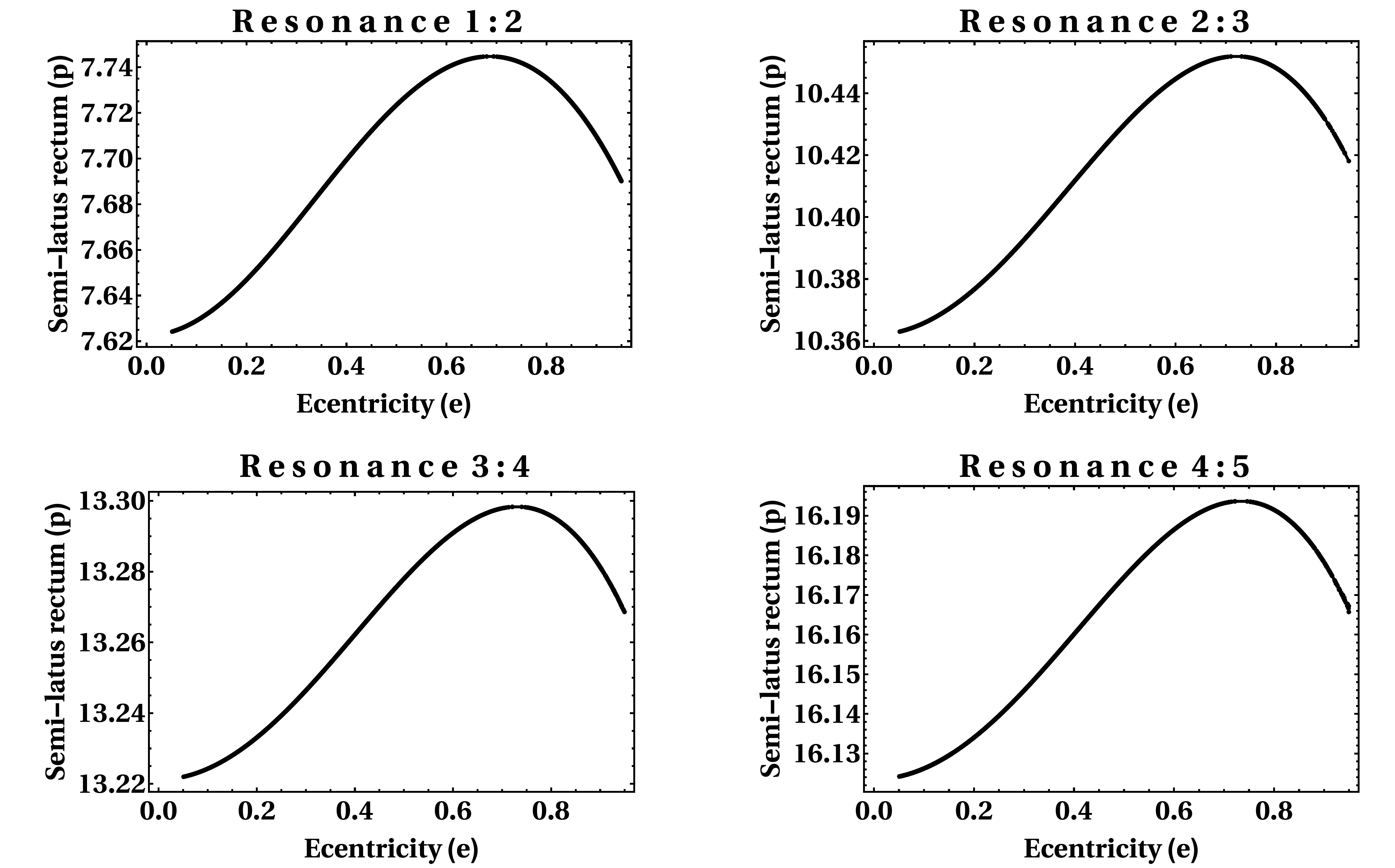}%
}\hfill
\subfloat[Resonant orbits are shown while the black hole has an angular momentum of $a=0.9M$. \label{fig:Resonance_08}]{%
  \includegraphics[height=6.2cm,width=.48\linewidth]{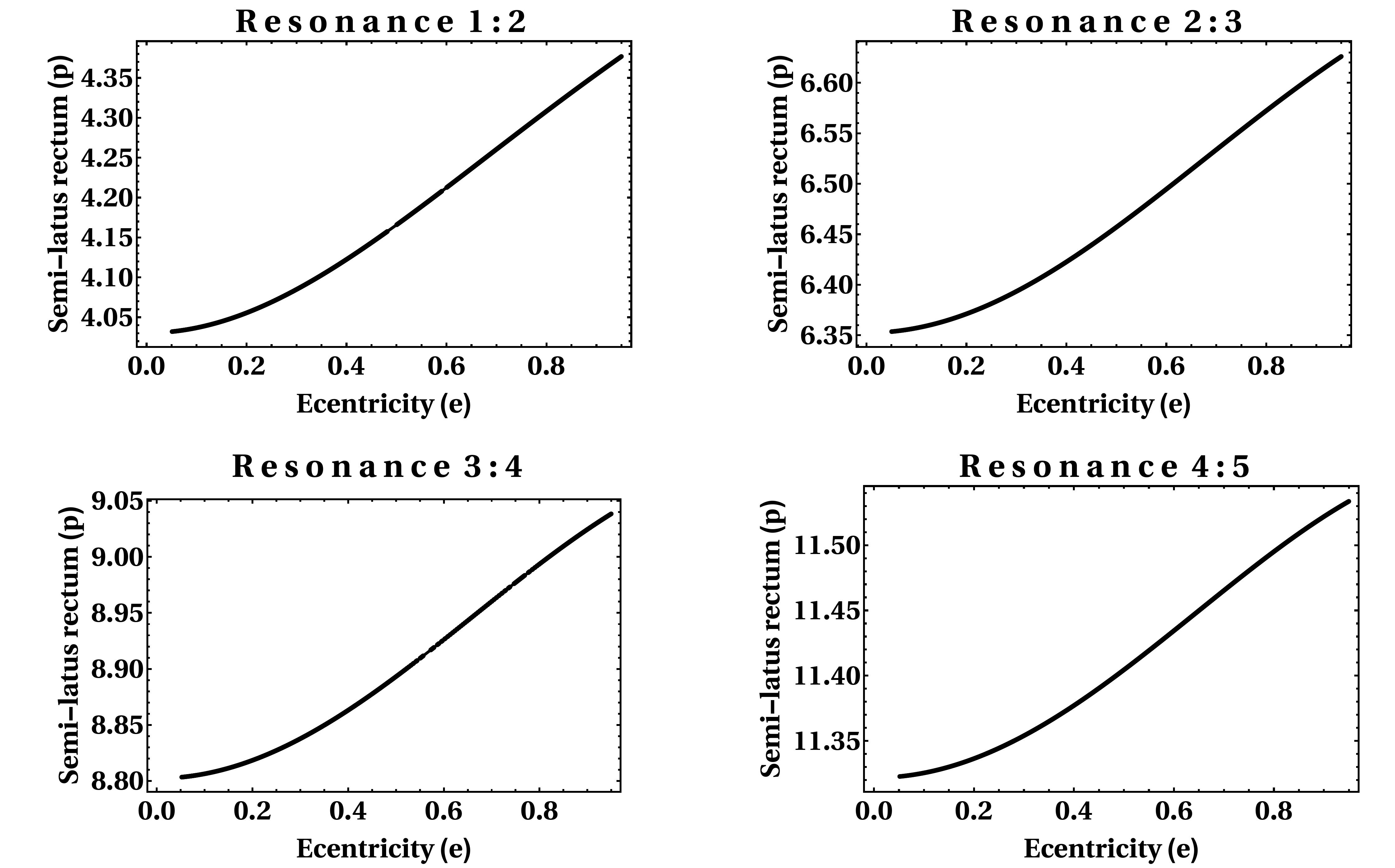}%
}
\caption{Above figure represents the $r\phi$ resonant orbits for particles with spin $S=10^{-4}M$. The influence of different angular momentum of the black hole is also depicted.}
\label{fig:Resonance_D}
\end{figure}

\begin{figure}[htp]
\subfloat[The resonant orbits for a spinning particle is shown in a black hole with angular momentum $a=0.099M$. \label{fig:Resonance_09}]{%
  \includegraphics[height=6.2cm,width=.48\linewidth]{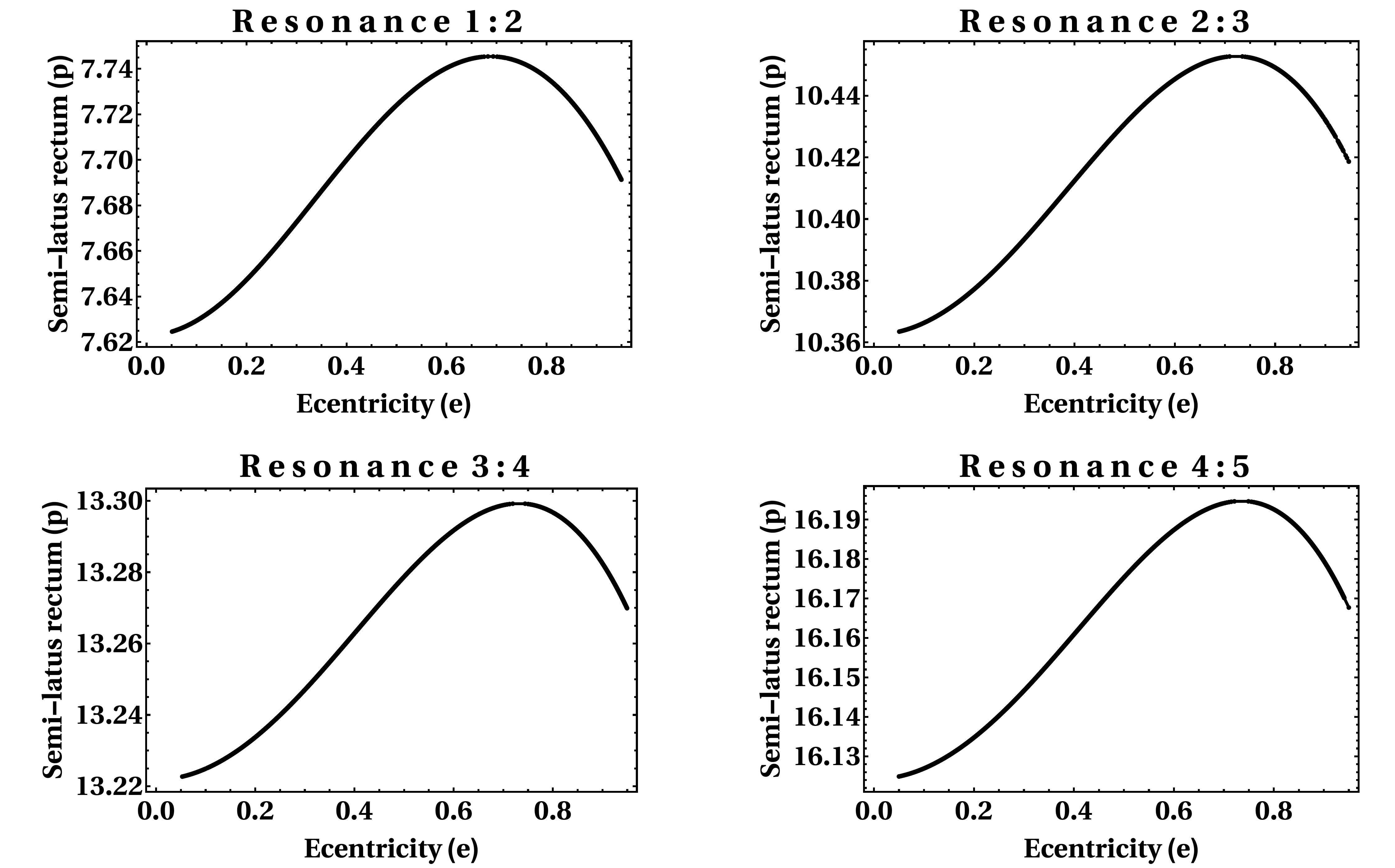}%
}\hfill
\subfloat[Resonant orbits are shown while the black hole has an angular momentum of $a=0.9M$. \label{fig:Resonance_10}]{%
  \includegraphics[height=6.2cm,width=.48\linewidth]{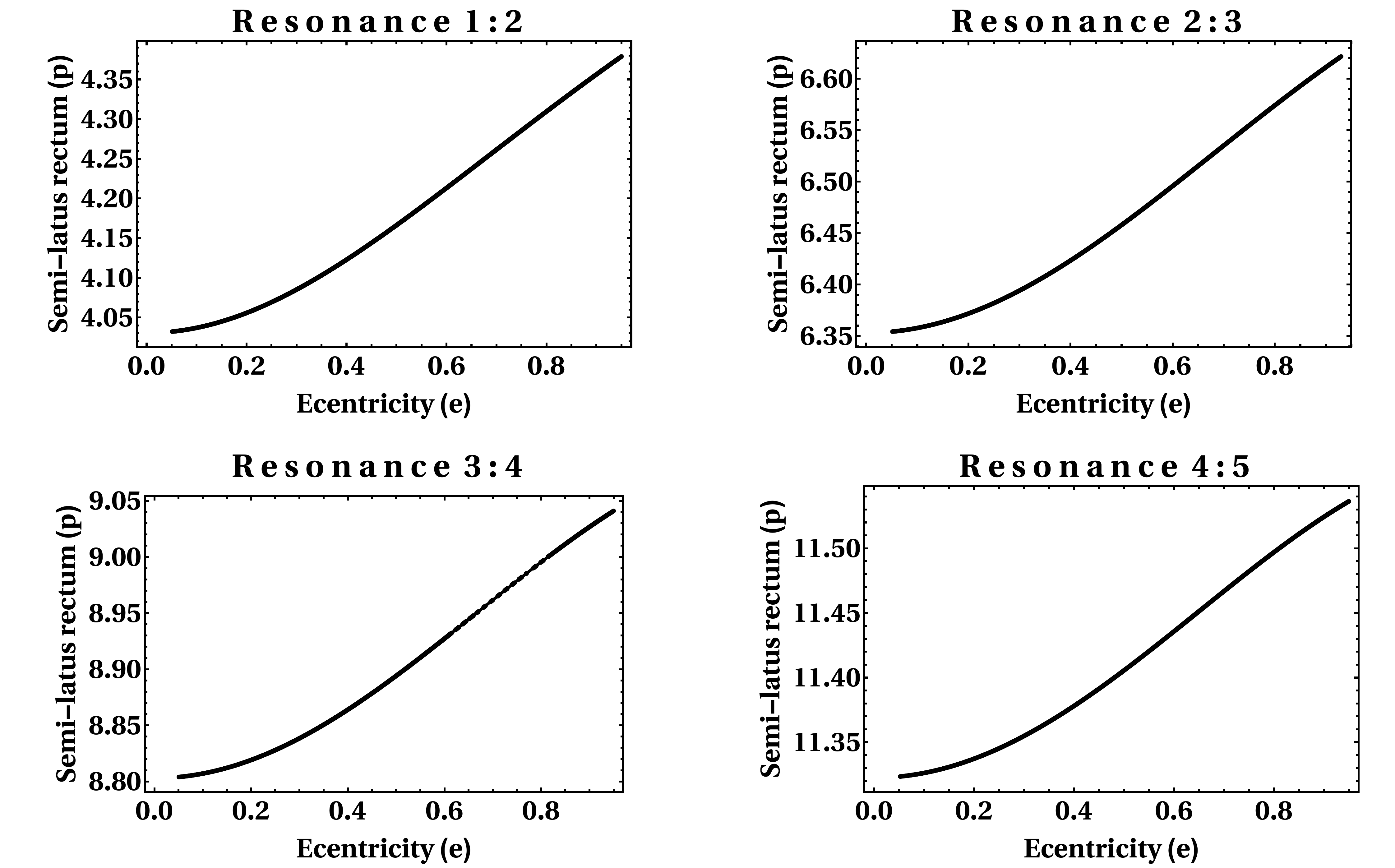}%
}
\caption{Above figure represents the $r\phi$ resonant orbits for particles with spin $S=-10^{-4}M$. The influence of different angular momentum of the black hole is also depicted.}
\label{fig:Resonance_E}
\end{figure}

%%%%%%%%%%%%%%%%%%%%%%%%%%%%%%%%%%%%%%%%%%%%%%%%%%%%%%%%%%%%%%%%%%%%%%%%%%%%%%%%%%%%%%%%%%%%
In this case, the radial potential within the linear spin approximation can be written in the following form
%%%%%%%%%%%%%%%%%%%%%%%%%%
\begin{eqnarray}
V_{\rm s}(r) &=& r^4 \biggl[E^2 r^4-(L^2_z-a^2E^2)r^2+2 M r(L_z-aE)^2-(r^2-2Mr+a^2)r^2\biggr]\nonumber\\
& & +S r^3 \left\{2 a^3 M E^2-4 a^2 M E J_z+2 a M J^2_z+6 a E^2 M r^2+2 E J_z r^3-6 E J_z M r^2\right\} \nonumber \\
&= & r^3 \biggl[r \biggl\{E^2 r^4-(L^2_z-a^2E^2)r^2+2 M r(L_z-aE)^2-(r^2-2Mr+a^2)r^2 \biggr\}+\nonumber\\
& & S \biggl\{2 a^3 M E^2-4 a^2 M E J_z+2 a M J^2_z+6 a E^2 M r^2+2 E J_z r^3-6 E J_z M r^2\biggr\}\biggr]\nonumber\\
& = & -\alpha_{\rm s} r^3(r-r_1)(r-r_2)(r-r_a)(r-r_p)(r-r_3),
\end{eqnarray}
%%%%%%%%%%%%%%%%%%%%%%%%%%%
and we now estimate each solutions by equating the left hand and right hand sides. Therefore, we arrive at the following set of equations
%%%%%%%%%%%%%%%%%%%%%%
\begin{enumerate}
\item  From coefficient of $r^5$, we have $\alpha_{\rm s}=1-E^2$.

\item From coefficient of $r^4$, we manage to have
%%%%%%%%%%%%%%%%%%%%%%%%
\begin{equation}
r_1+r_2+r_a+r_p+r_3=\dfrac{2M}{\alpha_{\rm s}}.
\label{eq:solve_E}
\end{equation}
%%%%%%%%%%%%%%%%%%%%%%%%
\item From coefficient of $r^3$, we have
%%%%%%%%%%%%%%%%%%%%%%%%%%
\begin{equation}
a^2(E^2-1)-J^2_z+2 E J_z S=-\alpha \left\{r_a r_p+r_1 r_2+(r_a+r_p)(r_1+r_2)+r_3(r_1+r_2+r_a+r_p)\right\}.
\label{eq:Jz_1}
\end{equation}
%%%%%%%%%%%%%%%%%%%%%%%%%%
\item From coefficient of $r^2$, we obtain
%%%%%%%%%%%%%%%%%%%%%%%%%%
\begin{eqnarray}
2 M a^2 E^2-4 a M E J_z+2 M J^2_z+6 a E^2 M S-6 E M S J_z &=&\alpha_{\rm s} \biggl\{r_a r_p r_3+(r_1+r_2)\biggl(r_a r_p+r_3 (r_a+r_p)\biggr).\nonumber\\
&&+r_1 r_2 (r_3+r_a+r_p)\biggr\}.
\label{eq:solve_r1}
\end{eqnarray}
%%%%%%%%%%%%%%%%%%%%%%%%%%%
\item From coefficient of $r$, we determine 
%%%%%%%%%%%%%%%%%%%%%%%%%%%
\begin{equation}
r_2 r_3 r_a r_p+r_1 r_3 r_a r_p+r_1 r_2 \left\{r_ar_p+r_3(r_a+r_p)\right\}=0.
\label{eq:solve_r3}
\end{equation}
%%%%%%%%%%%%%%%%%%%%%%%%%%%
\item From the coefficient of $r^0$, we get
%%%%%%%%%%%%%%%%%%%%%%%%%%%
\begin{equation}
(2M E^2 a^3-4MEJ_z a^2+2MaJ^2_z)S=r_1 r_2 r_3 r_a r_p \alpha_{\rm s}.
\label{eq:Jz_2}
\end{equation}
%%%%%%%%%%%%%%%%%%%%%%%%%%%
\end{enumerate}
%%%%%%%%%%%%%%%%%%%%%%
With $r_{+}=r_1+r_2$ and $r_{\times}=r_1 r_2$,  we can further compute $E$ from Eq. (\ref{eq:solve_E}),
\begin{equation}
E^2=1-\dfrac{2M}{r_{+}+r_a+r_p+r_3}.
\end{equation}
%%%%%%%%%%%%%%%%%%%%%%%%%%%
From Eq. (\ref{eq:solve_r3}), we can further compute $r_3$ as
%%%%%%%%%%%%%%%%%%%%%%%%%%%
\begin{equation}
r_3=-\dfrac{r_{\times} r_a r_p}{r_{\times}(r_a+r_p)+r_a r_p r_{+}},
\end{equation}
%%%%%%%%%%%%%%%%%%%%%%%%%%%
To obtain $J_z$, we can introduce the expression $X_{\pm}=2M a^2 \times $ Eq. (\ref{eq:Jz_1}) $\pm$ Eq. (\ref{eq:Jz_2}) and with the $+$ sign, we arrive at 
%%%%%%%%%%%%%%%%%%%%%%%%%%%
\begin{eqnarray}
J_z^2&=&{a(a-S)(r_++r_a+r_p+r_3)}^{-1}\Bigl\{2a^2M \Big[r_3 (r_a+r_p)+r_a r_p+r_{\times}+r_{+}(r_a+r_p+r_3)\Big] \nonumber \\
&& \hspace{5cm}-2M a^4-r_{\times}r_3 r_a r_p  + a^3 S (r_a+r_p+r_3+r_+-2M)\Bigr\}.
\end{eqnarray}
%%%%%%%%%%%%%%%%%%%%%%%%%%%
Along with the expressions $X_{-}$ and Eq. (\ref{eq:solve_r1}), we can numerically solve for both $r_1$ and $r_2$ for given values of $e$ and $p$. Therefore, whenever provided with $r_a$ and $r_p$, we can compute energy and momentum corresponding to a spinning particle confined on the equatorial plane of the Kerr black hole. The next task is to determine the integral given in Eq. (\ref{eq:primary}) and for a general Kerr spcatime, this is given in the Appendix. For convenience, we carry out the calculations within the linear framework of the spin parameter, while a general treatment shall be carried out elsewhere. In Fig. \ref{fig:Resonance_D}, and Fig. \ref{fig:Resonance_E} we have shown resonant orbits for different orders of resonance while the angular momentum of the Kerr black hole varies accordingly. For a particle with spin $S=10^{-4}M$, various resonant orbits for $a=0.099M$ and $a=0.9M$ are given in Fig.  \ref{fig:Resonance_07} and Fig. \ref{fig:Resonance_08} respectively. The studies related to spin $S=-10^{-4}M$ are shown in Fig. \ref{fig:Resonance_E} for different resonance and black hole's momentum.
%%%%%%%%%%%%%%%%%%%%%%%%%%%%
%%%%%%%%%%%%%%%%%%%%%%%%%%%%
%%%%%%%%%%%%%%%%%%%%%%%%%%%%
\section{Concluding remarks}\label{sec:remarks}
%%%%%%%%%%%%%%%%%%%%%%%%%%%%%
In the present article we have studied the trajectories of a spinning particle and discuss their orbital resonance in the Kerr background. We typically explore two particular events, first the resonance in-between small oscillation frequencies and second, $r\phi$ resonance on the equatorial plane of the Kerr black hole. In the case of a spinning particle, the off-equatorial trajectories are not easily derivable and can hardly be written in a convenient analytical form. Therefore, in the present context, we have only considered the $r\phi$ resonance and mostly concerned with locating the resonant orbits for different spin parameter. We confined our discussions for a pole-dipole particle, i.e, a particle with nonzero dipole moment while all the higher order moments are set to zero. In addition, the entire study is only valid for a linearized spin parameter and any contribution appears at $\mathcal{O}(S^2)$ and higher are ignored for convenience. The key findings of the present article can be summarized as follows. 

The first part of the paper deals with the nearly circular orbits hobbling around the equatorial plane, for an extended object around the Kerr black hole. In the presence of non-vanishing spin, the expressions for small oscillation frequencies would change and therefore, the locations of resonant orbits would also shift. For our study, we employed the analytical expressions for frequencies given by Hinderer et. al. in Ref. \cite{Hinderer:2013uwa} and obtained the locations of the resonant orbits for different angular momentum of the black hole. In the case of co-rotating orbits, we discuss all three resonance phenomena, i.e., $r\theta$, $r\phi$ and $\theta\phi$, given in Fig. \ref{fig:Resonance_01}, Fig. \ref{fig:Resonance_02}, and Fig. \ref{fig:Resonance_02a} respectively, for various spin parameters of the secondary object. On the other hand, the resonant orbits for counter-rotating case is demonstrated in Fig. \ref{fig:Resonance_Acnt}. The $r\theta$, $r\phi$, and $\phi\theta$ resonances are shown in Fig. \ref{fig:Resonance_01cnt}, Fig. \ref{fig:Resonance_02cnt}, and Fig. \ref{fig:Resonance_02acnt} respectively. In order to preserve the convention given in Eq. (\ref{eq:primary}), we address the resonance between $\theta$ and $\phi$ to be $\phi\theta$, rather than $\theta\phi$. As expected, within the linear spin approximation, the difference in location between geodesic and spinning particle is infinitesimal.

The second part deals with the $r\phi$ resonance considering the trajectories are completely confined on the equatorial plane. We locate the resonant orbits for different values of the black hole's momentum as well as the particle's spin. When the angular momentum of the black hole is small, the resonant orbits has a weak dependency on the eccentricity $e$. It is explicitly demonstrated in Fig. \ref{fig:Resonance_07} and Fig. \ref{fig:Resonance_09}. However, the $e$ dependency increases as one increases the value of angular momentum of the black hole as it is shown in Fig. \ref{fig:Resonance_08} and Fig. \ref{fig:Resonance_10}. The nature of resonant orbits are also largely effected by the black hole rotation as depicted in Fig. \ref{fig:Resonance_D} and Fig. \ref{fig:Resonance_E}. For a small or no rotation of the black hole (Fig. \ref{fig:Resonance_B}, Fig. \ref{fig:Resonance_07}, Fig. \ref{fig:Resonance_09}), the semi-latus rectum steadily increases with the eccentricity $e$ and attains a maximum value for a particular eccentricity say $e=e_{\rm max}$. Further increase in the eccentricity would result in the decrease of the semi-latus rectum $p$ as shown in Fig. \ref{fig:Resonance_B}, Fig. \ref{fig:Resonance_07}, and Fig. \ref{fig:Resonance_09}. On the other hand, for black holes with larger angular momentum, $p$ steadily increases with the eccentricity $e$ and attains the highest value as $e$ approaches unity. Furthermore, the orbits shift towards the event horizon as one either increases the momentum of the black hole or decreases the order of resonance.

Finally, it should be emphasized that the change in locations for different spin parameters within the linear approximation, is infinitesimally small and nearly indistinguishable. Therefore, it would be important to investigate for large spin parameters of the secondary object. However, the immediate follow up of the present work would be consider the $r\theta$ resonance which requires a numerical framework.
%%%%%%%%%%%%%%%%%%%%%%%%%%%%%%%%%%%%
\section{Acknowledgement}

The authors are indebted to Prof. Rajesh Kumble Nayak for many useful discussions on the present topic and for also suggesting helpful corrections about the manuscript. They are also thankful to the Center of Excellence in Space Sciences India (CESSI) for providing them with the computational facilities whenever required. Finally, the authors would like to thank the anonymous referee for some constructive comments and suggestions to improve the manuscript.

%%%%%%%%%%%%%%%%%%%%%%%%%%%%%%%%%%%%%%%%%%%%%%%%%%%%%%%%%%%%%%%%%%%%%%%%%%%%%%%%%%%%%%%%%%%%%%%%%%
\appendix
\section*{Appendix: Detail calculations to obtain the resonant orbits}
In the Kerr black hole, we obtain
%%%%%%%%%%%%%%%%%%%%%%%%%%%
\begin{equation}
\dfrac{\mathcal{U}^{3}}{\mathcal{U}^{1}}=\dfrac{d\phi}{dr}=\dfrac{r^2 \left\{J_z-(a+S)E\right\}+\dfrac{a P_s r^2}{\Delta}}{\biggl[V_{\rm s} (r)\biggr]^{1/2}},
\end{equation}
%%%%%%%%%%%%%%%%%%%%%%%%%%%
with $V_{\rm s}(r)=-\alpha_{\rm s} r^3(r-r_1)(r-r_2)(r-r_a)(r-r_p)(r-r_3)$. We can split the above integration into two parts, the first one is $\mathcal{G}$ and second one is $\mathcal{R}$ and these are given as
%%%%%%%%%%%%%%%%%%%%%%%%%%%
\begin{eqnarray}
\mathcal{G} &=&\dfrac{r^2 \left\{J_z-(a+S)E\right\}}{\biggl[V_{\rm s} (r)\biggr]^{1/2}}=\dfrac{r\left\{J_z-(a+S)E\right\}}{\biggl[-\alpha_{\rm s} (r-0)(r-r_1)(r-r_2)(r-r_a)(r-r_p)(r-r_3)\biggr]^{1/2}}, \nonumber \\
\mathcal{R} &=&\dfrac{a r^2 P_s}{\Delta \biggl[V_{\rm s} (r)\biggr]^{1/2}}=\dfrac{a \biggl\{E r^3+a r (a+S-J_z)+M S(a-J_z)\biggr\}}{(r-r^{+}_{\rm H})(r-r^{-}_{\rm H})\biggl[-\alpha_{\rm s} (r-0)(r-r_1)(r-r_2)(r-r_a)(r-r_p)(r-r_3)\biggr]^{1/2}}, \nonumber \\
\end{eqnarray}
%%%%%%%%%%%%%%%%%%%%%%%%%%%
where $r^{+}_{\rm H}$ and $r^{-}_{\rm H}$ are given as the outer and inner horizon respectively. Furthermore, with the substitution \cite{Kunst:2015hsa}
%%%%%%%%%%%%%%%%%%%%%%%%%%
\begin{equation}
r=\dfrac{(r_a+r_p)+x(r_a-r_p)}{2},
\end{equation}
%%%%%%%%%%%%%%%%%%%%%%%%%%
we arrive at the following expression
%%%%%%%%%%%%%%%%%%%%%%%%%%%
\begin{equation}
\dfrac{d\phi}{dx}=\dfrac{r_a-r_p}{2}\dfrac{d\phi}{dr}=\dfrac{r_a-r_p}{2}(\mathcal{G}_{x}+\mathcal{R}_x).
\end{equation}
%%%%%%%%%%%%%%%%%%%%%%%%%%%
The $\mathcal{G}$ becomes $\mathcal{G}_{x}$ and given as
%%%%%%%%%%%%%%%%%%%%%%%%%%%
\begin{equation}
\mathcal{G}_{x}=\dfrac{r_x\left\{J_z-(a+S)E\right\}}{\biggl[-\alpha_{\rm s} (r_x-0)(r_x-r_1)(r_x-r_2)(r_x-r_a)(r_x-r_p)(r_x-r_3)\biggr]^{1/2}}.
\end{equation}
%%%%%%%%%%%%%%%%%%%%%%%%%%%
By using the fact,
%%%%%%%%%%%%%%%%%%%%%%%%%%%
\begin{equation}
r_x-r_i=\dfrac{r_a+r_p-2 r_i}{2}\left(1+K_{i}x\right),
\end{equation}
%%%%%%%%%%%%%%%%%%%%%%%%%%%
with $K_{i}=\dfrac{r_a-r_p}{r_a+r_p-2 r_i}$ where $i$ runs from $0$ to $3$ and $r_{0}=0$, the above equation reads as
%%%%%%%%%%%%%%%%%%%%%%%%%%%
\begin{eqnarray}
\mathcal{G}_{x} &=&\dfrac{1}{2}{(r_a+r_p)}(1+K_0 x)\left\{J_z-(a+S)E\right\}\biggl\{2^{-6}\alpha_{\rm s}(1-x^2)(1+K_0 x)(1+K_1 x)(1+K_2 x)(1+K_3 x)\nonumber \\
& & \hspace{4cm}(r_a+r_p)(r_a-r_p)^2(r_a+r_p-2 r_1)(r_a+r_p-2r_2)(r_a+r_p-2r_3) \biggr\}^{-1/2},\nonumber \\
 &= &\dfrac{N_{\rm k}(1+K_0 x)\left\{J_z-(a+S)E\right\}}{\sqrt{\biggl[(1-x^2)(1+K_0 x)(1+K_1 x)(1+K_2 x)(1+K_3 x)\biggr]}}.\nonumber \\
\end{eqnarray}
%%%%%%%%%%%%%%%%%%%%%%%%%%%
In the above, we assume
%%%%%%%%%%%%%%%%%%%%%%%%%%%
\begin{equation}
N_{\rm k} =\dfrac{4(r_a+r_p)}{\sqrt{\biggl[\alpha(r_a+r_p)(r_a-r_p)^2(r_a+r_p-2 r_1)(r_a+r_p-2r_2)(r_a+r_p-2r_3)\biggr]}}=\dfrac{2 < N_{\rm k} > }{r_a-r_p},
\end{equation}
%%%%%%%%%%%%%%%%%%%%%%%%%%%
with
%%%%%%%%%%%%%%%%%%%%%%%%%%%
\begin{equation}
< N_{\rm k} > = \dfrac{2(r_a+r_p)}{\sqrt{\biggl[\alpha_{\rm s}(r_a+r_p)(r_a+r_p-2 r_1)(r_a+r_p-2r_2)(r_a+r_p-2r_3)\biggr]}}.
\end{equation}
%%%%%%%%%%%%%%%%%%%%%%%%%%%
Therefore, we can finally write
%%%%%%%%%%%%%%%%%%%%%%%%%%%
\begin{equation}
\mathcal{G}_{x}=\dfrac{2 < N_{\rm k} >}{r_a-r_p}\dfrac{(1+K_0 x)\left\{J_z-(a+S)E\right\}}{\sqrt{\biggl[(1-x^2)(1+K_0 x)(1+K_1 x)(1+K_2 x)(1+K_3 x)\biggr]}}.
\end{equation}
%%%%%%%%%%%%%%%%%%%%%%%%%%%
The other part can be written as
%%%%%%%%%%%%%%%%%%%%%%%%%%%
\begin{eqnarray}
\mathcal{R}_x &= &\dfrac{a \biggl\{E r_x^3+a r_x (a+S-J_z)+M S(a-J_z)\biggr\}}{(r_x-r^{+}_{\rm H})(r_x-r^{-}_{\rm H})\biggl[-\alpha_{\rm s} (r_x-0)(r_x-r_1)(r_x-r_2)(r_x-r_a)(r_x-r_p)(r_x-r_3)\biggr]^{1/2}},\nonumber\\
&=&\dfrac{a}{\Delta}{Er^3_x+a(a+S-J_z)r_x+MS(a-J_z)}\biggl\{2^{-6}\alpha_{\rm s}(1-x^2)(1+K_0 x)(1+K_1 x)(1+K_2 x)(1+K_3 x)\nonumber \\
& & \hspace{4cm} (r_a+r_p)(r_a-r_p)^2(r_a+r_p-2 r_1)(r_a+r_p-2r_2)(r_a+r_p-2r_3)\biggr\}^{-1/2},\nonumber\\
&=& \dfrac{2}{r_a-r_p}\dfrac{4a \times 4}{(r_a+r_p-2r^{+}_{\rm H})(r_a+r_p-2r^{-}_{\rm H})(1+K^{+}_{\rm H}x)(1+K^{-}_{\rm H}x)} \times \nonumber \\
 &  & \hspace{4cm}  \dfrac{1}{\sqrt{\alpha_{\rm s} (r_a+r_p) (r_a+r_p-2 r_1)(r_a+r_p-2r_2)(r_a+r_p-2r_3)}}\times\nonumber\\
& & \hspace{3cm}  \dfrac{Er^3_x+a(a+S-J_z)r_x+MS(a-J_z)}{(1+K^{+}_{\rm H}x)(1+K^{-}_{\rm H}x)\sqrt{(1-x^2)(1+K_0 x)(1+K_1 x)(1+K_2 x)(1+K_3 x)}}, \nonumber \\
&=& \dfrac{2 < \overline{N_{\rm k}} >}{r_a-r_p} \dfrac{E (1-K_0 x)^3 (r_a+r_p)^3+4a(a+S-J_z)(r_a+r_p) (1-K_0 x)+8MS(a-J_z)}{8(1+K^{+}_{\rm H}x)(1+K^{-}_{\rm H}x)\sqrt{(1-x^2)(1+K_0 x)(1+K_1 x)(1+K_2 x)(1+K_3 x)}},
\end{eqnarray}
%%%%%%%%%%%%%%%%%%%%%%%%%%%
where we assume
%%%%%%%%%%%%%%%%%%%%%%%%%%%
\begin{eqnarray}
< \overline{N_{\rm k}} > &=& \dfrac{4a \times 4}{(r_a+r_p-2r^{+}_{\rm H})(r_a+r_p-2r^{-}_{\rm H})(1+K^{+}_{\rm H}x)(1+K^{-}_{\rm H}x)} \nonumber \\ 
& & \times \dfrac{Er^3_x+a(a+S-J_z)r_x+MS(a-J_z)}{(1+K^{+}_{\rm H}x)(1+K^{-}_{\rm H}x)\sqrt{(1-x^2)(1+K_0 x)(1+K_1 x)(1+K_2 x)(1+K_3 x)}}.
\end{eqnarray}
%%%%%%%%%%%%%%%%%%%%%%%%%%%
The final equation can now be written as
%%%%%%%%%%%%%%%%%%%%%%%%%%%
\begin{eqnarray}
\dfrac{d\phi}{dx} &=&< N_{\rm k} >\dfrac{(1+K_0 x)\left\{J_z-(a+S)E\right\}}{\sqrt{\biggl[(1-x^2)(1+K_0 x)(1+K_1 x)(1+K_2 x)(1+K_3 x)\biggr]}}+\nonumber\\
&& < \overline{N_{\rm k}} > \dfrac{E (1-K_0 x)^3 (r_a+r_p)^3+4a(a+S-J_z)(r_a+r_p) (1-K_0 x)+8MS(a-J_z)}{8(1+K^{+}_{\rm H}x)(1+K^{-}_{\rm H}x)\sqrt{(1-x^2)(1+K_0 x)(1+K_1 x)(1+K_2 x)(1+K_3 x)}}.
\end{eqnarray}
%%%%%%%%%%%%%%%%%%%%%%%%%%%
By employing the fact, $\sqrt{(1+K_0 x)(1+K_1 x)}=1+\dfrac{(K_0+K_1)}{2}x=1+\overline{K}_1x$ and $\sqrt{(1+K_2 x)(1+K_3 x)}=1+\overline{K}_2 x$, the above equation can be written as
%%%%%%%%%%%%%%%%%%%%%%%%%%%
\begin{eqnarray}
\dfrac{d\phi}{dx} &=& < N_{\rm k} >\dfrac{(1+K_0 x)\left\{J_z-(a+S)E\right\}}{\sqrt{1-x^2}(1+\overline{K}_1x)(1+\overline{K}_2x)}+\nonumber\\
&& < \overline{N_{\rm k}} > \dfrac{E (1-K_0 x)^3 (r_a+r_p)^3+4a(a+S-J_z)(r_a+r_p) (1-K_0 x)+8MS(a-J_z)}{8\sqrt{(1-x^2)}(1+K^{+}_{\rm H}x)(1+K^{-}_{\rm H}x)(1+\overline{K}_1x)(1+\overline{K}_2x)}.
\end{eqnarray}
%%%%%%%%%%%%%%%%%%%%%%%%%%%
We further write
%%%%%%%%%%%%%%%%%%%%%%%%%%%
\begin{equation}
\dfrac{1}{(1+\overline{K}_1x)(1+\overline{K}_2x)}=\dfrac{1}{\overline{K}_2-\overline{K}_1}\left\{\dfrac{\overline{K}_2}{1+\overline{K}_2 x}-\dfrac{\overline{K}_1}{1+\overline{K}_1 x}\right\},
\end{equation}
%%%%%%%%%%%%%%%%%%%%%%%%%%%
and also
%%%%%%%%%%%%%%%%%%%%%%%%%%%
\begin{equation}
\dfrac{1}{(1+K^{+}_{\rm H}x)(1+K^{-}_{\rm H}x)(1+\overline{K}_1x)(1+\overline{K}_2x)}=\dfrac{A}{1+K^{+}_{\rm H}x}+\dfrac{B}{1+K^{-}_{\rm H}x}+\dfrac{C}{1+\overline{K}_1x}+\dfrac{D}{1+\overline{K}_2x},
\end{equation}
%%%%%%%%%%%%%%%%%%%%%%%%%%%
with $A$, $B$, $C$ and $D$ has the following expressions
%%%%%%%%%%%%%%%%%%%%%%%%%%%
\begin{eqnarray}
A &=& \dfrac{(K^{+}_{\rm H})^3}{(K^{+}_{\rm H}-K^{-}_{\rm H})(K^{+}_{\rm H}-\overline{K}_1)(K^{+}_{\rm H}-\overline{K}_2)}, \qquad B = \dfrac{(K^{-}_{\rm H})^3}{(K^{-}_{\rm H}-K^{+}_{\rm H})(K^{-}_{\rm H}-\overline{K}_1)(K^{-}_{\rm H}-\overline{K}_2)}, \nonumber \\
C &=& \dfrac{(\overline{K}_1)^3}{(\overline{K}_1-K^{+}_{\rm H})(\overline{K}_1-K^{-}_{\rm H})(\overline{K}_1-\overline{K}_2)}, \qquad D = \dfrac{(\overline{K}_2)^3}{(\overline{K}_2-K^{+}_{\rm H})(\overline{K}_2-K^{-}_{\rm H})(\overline{K}_2-\overline{K}_1)}.
\end{eqnarray}
%%%%%%%%%%%%%%%%%%%%%%%%%%%
With all the above machinery employed, we arrive at the following expression
%%%%%%%%%%%%%%%%%%%%%%%%%%%
\begin{eqnarray}
\dfrac{d\phi}{dx} &=&< N_{\rm k} > \dfrac{\left\{J_z-(a+S)E\right\}}{\overline{K}_2-\overline{K}_1} \biggl\{\dfrac{\overline{K}_2 (1+K_0 x)}{\sqrt{1-x^2}(1+\overline{K}_2 x)}-\dfrac{\overline{K}_1 (1+K_0 x)}{\sqrt{1-x^2}(1+\overline{K}_1 x)}\biggr \}+\nonumber\\
&&  \dfrac{<\overline{N_{\rm k}}>}{8} \biggl\{E  (r_a+r_p)^3 \biggl( \dfrac{A(1-K_0 x)^3}{\sqrt{(1-x^2)}(1+K^{+}_{\rm H}x)} +\dfrac{B(1-K_0 x)^3}{\sqrt{(1-x^2)}(1+K^{-}_{\rm H}x)}+\nonumber\\
&&\dfrac{C(1-K_0 x)^3}{\sqrt{(1-x^2)}(1+\overline{K}_1x)}+\dfrac{D(1-K_0 x)^3}{\sqrt{(1-x^2)}(1+\overline{K}_2x)}\biggr) \biggr\}+\dfrac{<\overline{N_{\rm k}}> a (a+S-J_z)(r_a+r_p)}{2}\biggl\{ \nonumber\\
& &  \dfrac{A(1-K_0 x)}{\sqrt{(1-x^2)}(1+K^{+}_{\rm H}x)} +\dfrac{B(1-K_0 x)}{\sqrt{(1-x^2)}(1+K^{-}_{\rm H}x)}+\dfrac{C(1-K_0 x)}{\sqrt{(1-x^2)}(1+\overline{K}_1x)}+\dfrac{D(1-K_0 x)}{\sqrt{(1-x^2)}(1+\overline{K}_2x)}\biggr \} \nonumber \\
&& + M S (a-J_z)<\overline{N_{\rm k}}> \biggl\{\dfrac{A}{\sqrt{(1-x^2)}(1+K^{+}_{\rm H}x)} +\dfrac{B}{\sqrt{(1-x^2)}(1+K^{-}_{\rm H}x)}+ \nonumber \\
& & \dfrac{C}{\sqrt{(1-x^2)}(1+\overline{K}_1x)}+\dfrac{D}{\sqrt{(1-x^2)}(1+\overline{K}_2x)}\biggr\}.
\end{eqnarray}
%%%%%%%%%%%%%%%%%%%%%%%%%%%
Now we have to do the calculation $\bigintsss_{-1}^{1}{\biggl(\dfrac{d\phi}{dx}\biggr)dx}$ and to do so, we have used the following equations
%%%%%%%%%%%%%%%%%%%%%%%%%%%
\begin{eqnarray}
I_1(K)&=&\int^{1}_{-1}\dfrac{dx}{\sqrt{1-x^2}(1+K x)}=\dfrac{\pi}{\sqrt{1-K^2}},\nonumber\\ 
I_2(K)&=&\int^{1}_{-1}\dfrac{(1+K_0 x)dx}{\sqrt{1-x^2}(1+K x)}=\dfrac{\pi}{K}\left\{\dfrac{K-K_0}{\sqrt{1-K^2}}+K_0\right\}, \nonumber \\
I_3(K)&=&\int^{1}_{-1}\dfrac{(1+K_0 x)^3dx}{\sqrt{1-x^2}(1+K x)}=\dfrac{\pi}{K^3}\left\{\dfrac{(K-K_0)^3}{\sqrt{1-K^2}}-\left[(K-K_0)^3-\dfrac{K^2(K^3_0+2K)}{2}\right]\right\}.
\end{eqnarray}
%%%%%%%%%%%%%%%%%%%%%%%%%%% 
With the above informations, we can write the final equation as follows
%%%%%%%%%%%%%%%%%%%%%%%%%%%%
\begin{eqnarray}
\int^{1}_{-1}\dfrac{d\phi}{dx}dx= &=&< N_{\rm k} > \dfrac{\left\{J_z-(a+S)E\right\}}{\overline{K}_2-\overline{K}_1} \biggl\{\overline{K}_2 I_2(\overline{K}_2)-\overline{K}_1 I_2(\overline{K}_1)\biggr\}+\nonumber\\
&&  \dfrac{<\overline{N_{\rm k}}>}{8} \biggl\{E  (r_a+r_p)^3 \biggl[ A I_3(K^{+}_{\rm H})+B I_3(K^{-}_{\rm H})+C I_3(\overline{K}_1)+D I_3(\overline{K}_2) \biggr] \biggr\}+ \nonumber \\
& & \dfrac{<\overline{N_{\rm k}}> a (a+S-J_z)(r_a+r_p)}{2}\biggl\{A I_2(K^{+}_{\rm H})+B I_2(K^{-}_{\rm H})+C I_2(\overline{K}_1)+D I_2(\overline{K}_2) \biggr \} \nonumber \\
&& + M S (a-J_z)<\overline{N_{\rm k}}> \biggl \{A I_1(K^{+}_{\rm H})+B I_1(K^{-}_{\rm H})+C I_1(\overline{K}_1)+D I_1(\overline{K}_2)\biggr \}.
\end{eqnarray}
%%%%%%%%%%%%%%%%%%%%%%%%%%%%

%%%%%%%%%%%%%%%%%%%%%%%%%%%%%%%%%%%%%%%%%%%%%%%%%%%%%%%%%%%%%%%%%%%%%%%%%%%%%%%%%%%%%%
\bibliographystyle{utphys1}
\bibliography{References}
%%%%%%%%%%%%%%%%%%%%%%%%%%%%%%%%%%%%%%%%%%%%%%%%%%%%%%%%%%%
%%%%%%%%%%%%%%%%%%%%%%%%%%%%%%%%%%%%%%%%%%%%%%%%%%%%%%%%%%%%%%%%%%%%%%%%%%%%%%%%%%%%%%%%%%%%%%%%%%

\end{document}